\documentclass[12pt]{article}
\usepackage{amssymb, amsmath}
\usepackage{ulem}
\usepackage{amsfonts}
\usepackage{graphicx}
\usepackage{setspace} 
\usepackage{rotating}
\usepackage[utf8]{inputenc}
\usepackage[]{natbib}
\usepackage{array,epsfig,fancyheadings,rotating}
\usepackage{caption}
\usepackage{subcaption}
\usepackage{amsmath}
\usepackage{hyperref}
\hypersetup{
  colorlinks=true,
  linkcolor=blue,
  urlcolor=blue,
  citecolor=blue,
}
\usepackage{amsmath}
\usepackage{booktabs}
\usepackage{multirow}
\usepackage{threeparttable}
\usepackage{adjustbox}

\usepackage{amssymb}
\usepackage{amsfonts}

\usepackage{xcolor}
\usepackage{amsthm}
\usepackage{lscape}
\usepackage{afterpage}
\usepackage{enumitem}

\usepackage{algorithm}
\usepackage{algpseudocode}
\usepackage{authblk}

\theoremstyle{definition}

\newcommand{\bh}{\mathbf{h}}

\newcommand{\bbP}{\mathbb{P}}

\newcommand{\bbR}{\mathbb{R}}

\newcommand{\E}{\mathbb{E}}

\title{Covariate-Adjusted Functional Principal Components Analysis for Modeling Hazard Rates of Physical Activity in the US Population}
\author[1]{Md Rokibul Hasan}
\author[1,$\ast$]{Pratim Guha Niyogi}

\affil[1]{Department of Data Science, University of Mississippi Medical Center}
\affil[$\ast$]{Corresponding author: pguhaniyogi@umc.edu}

\begin{document}
\maketitle

\begin{abstract}
Physical activity plays a vital role in human health. Its entire distribution differs among people. Commonly used summary measures cannot describe this distributional pattern. We present a distribution-based analytical approach to describe physical activity by modeling individual-level activity--intensity patterns through hazard functions derived from wrist-worn accelerometer data.

We analyzed minute-level Monitor-Independent Movement Summary (MIMS) data of 4,297 adults with seven continuous days of device wear from the 2011--2012 National Health and Nutrition Examination Survey (NHANES). We derived a nonparametric activity--intensity hazard using a survival-based approach for each individual on a common intensity grid, treating both the hazard curves from MIMS and their log-transformed MIMS as functional objects. We used functional principal component analysis (FPCA) on both scales of MIMS to characterize dominant modes of variation in activity--intensity distributions.

Group-wise mean hazard functions showed little difference at lower intensity levels, while we observed a substantial difference at higher intensity levels. Specifically, women, older adults, individuals with obesity, and hypertensive participants demonstrated more hazards at higher activity levels. The first functional component (PC1 corresponding to eigenvalue $\lambda_1$) explained around 60--70\% of the total variability. The dominant functional variability (PC1) was observed in low-to-moderate intensity regions on the original MIMS scales; however, on the log-transformed scale of MIMS showed the first dominant variability in moderate-to-high intensity ranges. Results from stratified FPCA showed primary modes of variation (PC1) have almost the same similarity across different groups, whereas higher-order components covered subgroup-specific variability.

Our results demonstrate that hazard-based functional representations for capturing differences in physical activity intensity distributions across individuals offer a flexible and interpretable way to characterize heterogeneity. This approach works better than mean-based summaries and supports principled comparisons of physical activity patterns across population subgroups.
\end{abstract}


\section{Introduction}

Physical activity is a cornerstone of a healthy and long life. It is defined as any movement produced by the body’s muscles that requires energy expenditure \citep{caspersen1985physical, piggin2020physical}. Physical activity helps people live longer \citep{ludlow2011physical} and plays a crucial role in reducing the overall burden of disease \citep{BAUMAN20046}. It is a major modifiable risk factor for many cardiovascular and chronic diseases \citep{WARBURTON2016495}. Moreover, physical activity, together with proper nutrition, promotes functional ability and supports the development of a strong immune system. Physically less active individuals are more prone to weaker immune responses, which makes them more vulnerable to diseases \citep{shao2021physical}. With recent advances in technology, accelerometers and other wearable devices now enable continuous monitoring of a wide range of physical activity behaviors under free-living conditions. These devices provide objective, high-resolution measurements that capture daily movement patterns across the entire day, thereby characterizing an individual’s 24-hour activity profile. Such detailed and continuous data offer new opportunities to study physical activity in real-world settings and better to understand its relationship with health and disease outcomes \citep{troiano2014evolution, van2015objective, migueles2017accelerometer, doherty2017large}. The pattern of physical activity across different times in a day is also an important determinant of health. Low-activity over the whole day is associated with health hazards such as worse metabolic health. More time with sedentary behavior results in worse health outcomes compared with low-intensity physical activity (LIPA) and moderate-to-vigorous physical activity (MVPA) \citep{Rossen02042020}. More importantly, the pattern of physical activity does not work the same way for each type of respondent \citep{jefferis2016does}.
\par

Most importantly, physical activity is not a uniform phenomenon; rather, it varies across the population. The activity pattern varies in terms of amount, intensity, and timing, varied by biological, demographic, and clinical characteristics. Factors such as age, sex, body composition, and cardiometabolic health status impact levels of physical activity and the pattern in which activity is distributed throughout the day \citep{hyde2013enhancing,jefferis2016does,guthold2018worldwide,Rossen02042020}. As a consequence, population averages may not reveal meaningful differences in activity patterns across subgroups \citep{doherty2017large,migueles2017accelerometer}.
So, understanding differences in physical activity patterns across different demographic and health-related groups is very important for accurately capturing movement behaviors and their health implications. 

Physical activity patterns change over the human life course. Among most people, it is higher in early ages and declines with aging \citep{hyde2013enhancing, mohd2021review}. Some studies found different patterns over the life span, too. \citet{malina1996tracking} showed that physical inactivity in childhood often tracks into inactive adulthood, whereas moderate activity levels in childhood tend to persist as moderately active behavior in adulthood. There are differences between men and women in terms of physical activity, though there are some inconsistencies. Studies found girls and women to be more active than boys and men \citep{guthold2018worldwide}. Sometimes discrimination against women seems to be a factor in higher physical activity among women \citep{moreno2022gender, altenburg2022physical}. \citet{trost1996gender} also reported contrasting findings, showing that boys and men tended to be more physically active, a result that is consistent with the observations of \citet{sallis1996ethnic} where men seemed more engaged in physical activity. 
\par
Obesity has turned out to be a global burden in public health \citep{jakicic2011obesity, ncd2016trends}. Higher categories of body mass index (BMI), particularly obesity, limit individuals’ ability to be physically active, as they are more likely to experience fatigue and reduced tolerance to physical activity. \citep{ortega2019cardiorespiratory, kilincarslan2019effect, sagelv2021bidirectional}. It also makes people vulnerable to many diseases, which is another reason for inactivity \citep{germain2016physical}. People with obesity face challenges in spending their energy through physical movements \citep{henson2016sedentary}. It is like a vicious cycle that who have high BMI are less active, which causes more high BMI, that results into more physical activity \citep{hsu2011vicious}. 
\par
Physical activity reduces the mortality risks due to hypertension significantly \citep{rossi2012impact, vieira2024barriers}. Hypertensive individuals are prone to being less physically active. Physical inactivity is a proven risk factor for hypertension \citep{hasan2024analyzing}. Less physically active people show higher odds of getting hypertension. For improving vascular function and reducing arterial stiffness, physical activity can play a key role \citep{ewunie2022physical, sakellariou2021exercise}.  
\par
Most studies have studied the effect of physical activity on biological, demographic, and clinical characteristics, but have ignored the distribution of activity over time. In addition, as the time of day is fixed, levels of activity- sleep, sedentary behaviour, LIPA, and MVPA are correlated. If any individual spends more time in any patterns of activity that definitely affects another type of physical activity \citep{chastin2015combined, tremblay2016canadian}. Recent studies have increasingly moved beyond traditional mean-based analyses and adopted more informative distribution-based approaches for studying physical activity patterns. These methods enable researchers to characterize heterogeneity across the entire distribution of activity rather than focusing solely on average behavior. One prominent example is the quantile-based approach, which provides insight into how covariate effects may differ across different levels of physical activity \citep{ghosal2023distributional, mendez2025functional, niyogi2026quantifying}. In this direction, \citet{chiou2009modeling} proposed a functional data approach for modeling age-specific hazard rates, enabling the characterization of temporal variation in mortality patterns and improving mortality forecasting. Recent studies \citep{niyogi2024hazard,guha_niyogi_2026_scalar_on_distribution} show the different distributional representation in the context of the subject with multiple sclerosis.

In this work, we adopt a functional data analytic approach to characterize physical activity through subject-specific activity–intensity distributions, enabling the investigation of heterogeneity in activity patterns beyond traditional summary measures. Functional data analysis (FDA) \citep{crainiceanu_fda, ramsay2005functional, ferraty2006nonparametric, horvath2012inference, wang2016functional, hsing2015theoretical, kokoszka2017introduction} provides a flexible framework for studying data that are naturally represented as functions rather than scalar measurements. By treating activity–intensity distributions as functional observations, FDA allows us to examine variation in physical activity patterns across the entire distribution and to assess how these patterns differ according to demographic and health-related characteristics. This study utilizes data from the enhanced Physical Activity Monitor (PAM) component of the National Health and Nutrition Examination Survey (NHANES) 2011–2012 cycle. NHANES is a nationally representative survey conducted by the National Center for Health Statistics (NCHS) of the Centers for Disease Control and Prevention (CDC) and combines interviews, physical examinations, and laboratory assessments to evaluate the health status of the U.S. population. In this cycle, NHANES reintroduced objective physical activity monitoring using wrist-worn ActiGraph GT3X+ accelerometers, which continuously recorded triaxial acceleration data at a sampling frequency of 80 Hz over a seven-day monitoring period. Unlike previous hip-worn devices, the wrist-worn protocol enabled the collection of 24-hour movement data, capturing both wake and sleep behaviors with high participant compliance. The enhanced accelerometry data are publicly available through the NHANES website (\url{https://wwwn.cdc.gov/nchs/nhanes/}) and have become an important resource for studying physical activity, sleep, circadian behavior, and health outcomes in the U.S. population.
This high temporal resolution of wrist-worn accelerometer data from NHANES 2011-2012 provides an opportunity to move beyond traditional summary measures of physical activity (e.g., total activity volume, average activity intensity, or time spent in predefined activity categories) and study the entire distribution of the activity profile. In particular, individuals with similar average activity levels may exhibit substantially different distributions of activity intensity throughout the day, potentially reflecting underlying differences in demographic characteristics, health status, and clinical conditions.

These considerations motivate the primary research question of this study: Do individuals with hypertension exhibit different physical activity patterns, as characterized by hazard-based functional representations of activity intensity, compared with individuals without hypertension? To address this question, we represent each individual's physical activity profile as a hazard function derived from the distribution of activity intensity, where the hazard characterizes the instantaneous rate of change in activity intensity. Treating these hazard functions as functional observations, we employ functional principal component analysis (FPCA) to investigate variation in physical activity patterns across age groups, sex, BMI categories, and hypertension status. This approach enables the characterization of distributional features of physical activity that may not be captured by conventional summary measures. To the best of our knowledge, this is the first study to examine physical activity using hazard-based functional representations constructed from nationally representative accelerometer data.

The remainder of this article is organized as follows. Section \ref{sec:data} describes the dataset and study variables. Section \ref{sec:model} presents the statistical methodology and modeling framework used in this research. The results are presented in Section \ref{sec:results}. Finally, Section \ref{sec:discussion} concludes with a discussion of the findings. Additional figures and supporting materials are provided in the Supplementary Material.

\section{Data description}
\label{sec:data}
We used publicly available data from the National Health and Nutrition Examination Survey (NHANES) for this study \citep{johnson2014national}. The dataset consists of high-resolution, continuously recorded wrist-worn accelerometry measurements from a large, nationally representative sample of individuals in the United States. Moreover, the survey also provides comprehensive data on the health and nutritional status of both adults and children, covering dietary intake, supplement use, and biomarker-based measures of nutrient status. Our focus was on the demographic, blood pressure-related information, and monitor independent movement summary (MIMS) values as an objective measure of physical movement and activity level collected from Wrist-worn accelerometry collected from the NHANES 2011-2012 dataset. Minute-level MIMS-unit acceleration measurements from the x-, y-, and z-axes are aggregated up and released as a variable in the minute summary. The minute-level MIMS-unit calculation was based on a universal summary metric and does not account for the ``valid/invalid'' QC flags. Participants were asked to wear the wrist accelerometer continuously for 24 hours per day over multiple consecutive days, including during sleep, over multiple consecutive days, following the study protocol. This continuous wear protocol ensures that daily activity and rest patterns are fully captured, allowing movement behaviors to be assessed over the entire 24-hour period. There are a total of 9756 individuals in the NHANES 2011-2012 study. To ensure the quality of the physical activity data, we only considered the subjects who had 7 distinct days of the week as good days and had the data regarding age, gender (not indicating the conflation of sex and gender), body mass index (BMI), systolic and diastolic blood pressure, and information about taking antihypertensive medication. Individuals are defined as hypertensive if reporting a diagnosis having an average systolic blood pressure of $\geq 130$ or an average diastolic blood pressure $\geq 80$ or taking any antihypertensive medication \citep{peng2023association}. Thus, overall we have 4448 subjects, where 51.06\% of them are female with an average age 45.4 (SD = 17.2), an average BMI 28.9 (SD = 7.05), and  45.70\% of them have hypertension. We group based on age group such as $18-35$ (33.52\%), $35-50$ (25.65\%), $50-65$ (26.37\%) and $65-79$ (14.45\%). BMI is classified as underweight (2.09\%) when BMI $\leq 18.5$, normal weight (29.63\%) when the BMI ranges from $18.5-25$, overweight (31.07\%) when BMI is within 25 and 30, obese (36.10\%) when the BMI $\geq 30$ \citep{consultation2000obesity}. Thus, we have 4297 subjects, where 50.5\% of them are female with an average age of 45.5 (SD = 17.2), an average BMI of 28.9 (SD = 7.06), and  46.80\% of them have hypertension. We categorize in groups based on age group such as $18-35$ (33.37\%), $35-50$ (25.41\%), $50-65$ (26.62\%) and $65-79$ (14.59\%). BMI is classified as underweight (2.07\%) when BMI $\leq 18.5$, normal weight (30.11\%) when the BMI ranges from $18.5-25$, overweight (31.28\%) when BMI is within 25 and 30, obese (36.54\%) when the BMI $\geq 30$ \citep{consultation2000obesity}.

\section{Modeling hazard function for PA}
\label{sec:model}
This research analyzes a full week's minute-level activity data, where the distribution of physical activity data is considered as a functional object. We ignore the time stamp of PA; rather, we are interested in the distribution of intensities using a hazard function. The hazard function models how common or rare different physical activity intensities are over the day.
This study evaluates the activity-intensity hazard function of each respondent's activity. The hazard function of activity counts represents the instantaneous rate at which activity counts exceed a given value. This conditional probability is particularly useful for studying the tail behavior of the distribution. A rapidly declining hazard for large values of activity count indicates that extremely high activity counts are increasingly rare. 
\par
Suppose $X_{ij}$ is the nonnegative random variable that represents the MIMS or log-transformed MIMS for the $i$-th subject at $j$-th time point for $i=1, \cdots, n$ and $j = 1, \cdots, m$. Further assume that $\{ X_{ij}: j = 1, \cdots, m_{i}\} \sim f_{i}$ and $X_{ij}$ are non-negative almost everywhere. Then we can define the hazard function evaluated at $x > 0$ as 
\begin{equation}
    h_{i}(x) = \lim_{\delta \rightarrow 0} \frac{1}{\delta}\bbP\{X_{i} \in [x, x+\delta] | X_{i} > x\}
\end{equation}
To capture the major mode of variation in the activity intensity through the hazard function, we use the functional principal component analysis (fPCA). We express the smoothed hazard function using the Kosambi–Karhunen–Lo\'eve (KKL) decomposition \citep{karhunen1946spektraltheorie, loeve1946functions}, which enables us to decompose the subject-specific hazard function into a systematic part corresponding to the overall mean function and a random part. 
\begin{equation}
\label{eq:fpca}
    h_{i}(x) = \mu(x) + \sum_{k=1}^{\infty}\xi_{ik}\phi_{k}(x)
\end{equation}
In Equation \eqref{eq:fpca}, $\mu(x)$ denotes the overall mean function, and $\phi_{k}(x)$ represents the $k$-th orthonormal eigen-function. The $\xi_{ik}$ are the functional principal component scores. These are the random variables $\xi_{ik} = \int (h_{i}(x) - \mu(x))\phi_{k}(x) dx$ with $\E\{\xi_{ik}\} = 0$ and $\E\{\xi_{ik}^{2}\} = \lambda_{k}$ where $\lambda_{k}$ is the eigen-value corresponding to the eigen-function $\phi_{k}$ for $k \geq 1$. Let $Z$ denote a categorical covariate defining $g$ distinct groups (e.g., gender, age group, hypertension status), so that $Z \in \{1, \ldots, g\}$. 
In the presence of $Z$, we consider the subject-specific conditional hazard function 
$h_{i}(x | Z = z) = \lim_{\delta \rightarrow 0} \frac{1}{\delta}\bbP\{X_{i} \in [x, x+\delta] | X_{i} > x, Z = z\}$ for $z \in \{1, \cdots, g\}$. Motivated by the model in Equation \eqref{eq:fpca}, we define a conditional version for the group-wise subject-specific hazard functions $h_{i}(x|Z = z)$ as 
\begin{equation}
\label{eq:haz}
    h_{i}(x|Z = z) = \mu(x|Z = z) + \sum_{k=1}^{\infty}\xi_{ikz}\phi_{k}(x|Z = z),
\end{equation}
for $z \in \{1, \cdots, g\}$, where $\mu(x|Z= z)$ and $\phi_{k}(x|Z = z)$ are the mean and eigen-functions as we defined before, along with the restriction that $Z = z$. Similarly, we define the functional principal component scores $\xi_{ikz} = \int \{h_{i}(x|Z = z) - \mu(x|Z = z)\}\phi_{k}(x| Z = z) dx$ with $\E\{\xi_{ikz}\} = 0$ and $\E\{\xi_{ikz}^{2}\} = \lambda_{kz}$. The model in Equation \eqref{eq:haz} is nonparametric; thus, we don't need to assume the shape of the hazard functions or the eigen-functions. The mean function $\mu(x|Z = z)$ is estimated for each group $z$. The eigen-function $\phi_{k}(x|Z = z)$ characterizes the main direction of the deviation of the hazard function for the group $z$. These random deviations are decomposed into two major terms, such as the eigen-functions and the scores, which vary according to the size of these deviations for each group. The subject-specific scores $\hat{\xi}_{ikz}$ indicate how each subject’s hazard curve aligns with the principal directions of variation. A set of ordered eigenfunctions $\hat{\phi}_k(x| Z = z)$, ranked by the amount of variance they explain through the associated eigenvalues $\hat{\lambda}_{kz}$, which quantify the variability captured by each score. 
\par
It is important to emphasize that, in practice, the hazard function is not directly observable from wearable accelerometry data. 
Instead, it must be estimated from the empirical distribution of activity counts. 
For subject $i$ with group membership $Z_i = z$, let $\{X_{ij}: j=1,\ldots,m_i\}$ denote the observed activity counts and let 
$x_{i(1)} < \cdots < x_{i(K_i)}$ be the distinct observed activity levels. 
Define the number of observations at level $x_{i(k)}$ as $d_{ik} = \sum\limits_{j=1}^{m_i} \mathbf{1}\{X_{ij} = x_{i(k)}\}$, and the corresponding risk set size as $R_{ik} = \sum\limits_{j=1}^{m_i} \mathbf{1}\{X_{ij} \ge x_{i(k)}\}$. We first construct the product-limit estimator of the survival function for the activity-intensity distribution, $\widehat{S}_i(x| Z = z) = \prod\limits_{k: x_{i(k)} \le x}\left(1 - \frac{d_{ik}}{Y_{ik}}\right)$ and then estimate the cumulative hazard via the Nelson-Aalen estimator, $\widehat{H}_i(x| Z = z) 
= \sum\limits_{k: x_{i(k)} \le x}\frac{d_{ik}}{Y_{ik}}$. Finally, the subject-specific hazard function is obtained by smoothing the increments of $\widehat{H}_i(\cdot)$ and differentiating the resulting smooth curve. 
Specifically, let $\widetilde{H}_i(x| Z = z)$ denote a smooth approximation to the step function $\widehat{H}_i(x| Z = z)$ obtained via kernel smoothing or spline-based regression. For example, one may define $\widetilde{H}_i(x) 
= \int K_\tau(x - u) \, d\widehat{H}_i(u| Z = z)$, 
where $K_\tau(\cdot) = \tau^{-1}K(\cdot/\tau)$ is a kernel function with bandwidth $\tau > 0$. 
The hazard function estimator is then given by $\widehat{h}_i(x| Z = z) = \frac{d}{dx}\widetilde{H}_i(x| Z = z)$. Under standard regularity conditions on the kernel and bandwidth selection, the resulting estimator $\widehat{h}_i(x| Z = z)$ is consistent and admits a bias-variance tradeoff controlled by $\tau$.
Equivalently, the cumulative hazard function may be regularized using spline-based methods, which provide a flexible yet stable framework for functional estimation. We approximate $\widehat{H}_i(x| Z = z)$ by a smooth spline function of the form $\widetilde{H}_i(x| Z = z) = \sum\limits_{\ell=1}^{L} \theta_{i\ell} B_\ell(x)$, 
where $\{B_\ell(\cdot)\}_{\ell=1}^{L}$ is a B-spline basis of degree $p$ defined over a suitably chosen knot sequence, and $\boldsymbol{\theta}_i = (\theta_{i1}, \ldots, \theta_{iL})^\top$ are spline coefficients. The coefficients are obtained by minimizing a penalized least squares criterion,
\begin{equation}
    \boldsymbol{\widehat{\theta}}_i 
= \arg\min_{\boldsymbol{\theta}}
\left\{
\int \left[ \widehat{H}_i(x) - \sum_{\ell=1}^{L} \theta_{\ell} B_\ell(x) \right]^2 w_i(x)\,dx
+ \lambda \int \left[ \frac{d^2}{dx^2} \sum_{\ell=1}^{L} \theta_{\ell} B_\ell(x) \right]^2 dx
\right\},
\end{equation}
where $w_i(x)$ is a weight function and $\lambda > 0$ is a smoothing parameter controlling the tradeoff between fidelity to the empirical cumulative hazard and smoothness of the estimate. 
The penalty on the second derivative enforces global smoothness and prevents overfitting to the discrete jumps of $\widehat{H}_i(x)$. Thus, the hazard function estimator is then obtained by differentiation, $\widehat{h}_i(x| Z = z) 
= \frac{d}{dx}\widetilde{H}_i(x)
= \sum_{\ell=1}^{L} \widehat{\theta}_{i\ell} B_\ell'(x)$, which remains a spline function of degree $p-1$. This representation ensures that $\widehat{h}_i(x| Z = z)$ is continuous (and sufficiently smooth depending on $p$), facilitating subsequent functional modeling and covariance estimation. Spline-based regularization is particularly advantageous for dense accelerometry data, where the empirical cumulative hazard may exhibit numerous small jumps due to high-frequency measurements. 
The penalized spline framework stabilizes these fluctuations while retaining the essential structural features of the activity-intensity distribution.
\par
\begin{figure}
    \centering
    \begin{subfigure}[t]{0.48\textwidth}
        \centering
        \includegraphics[width=\textwidth]{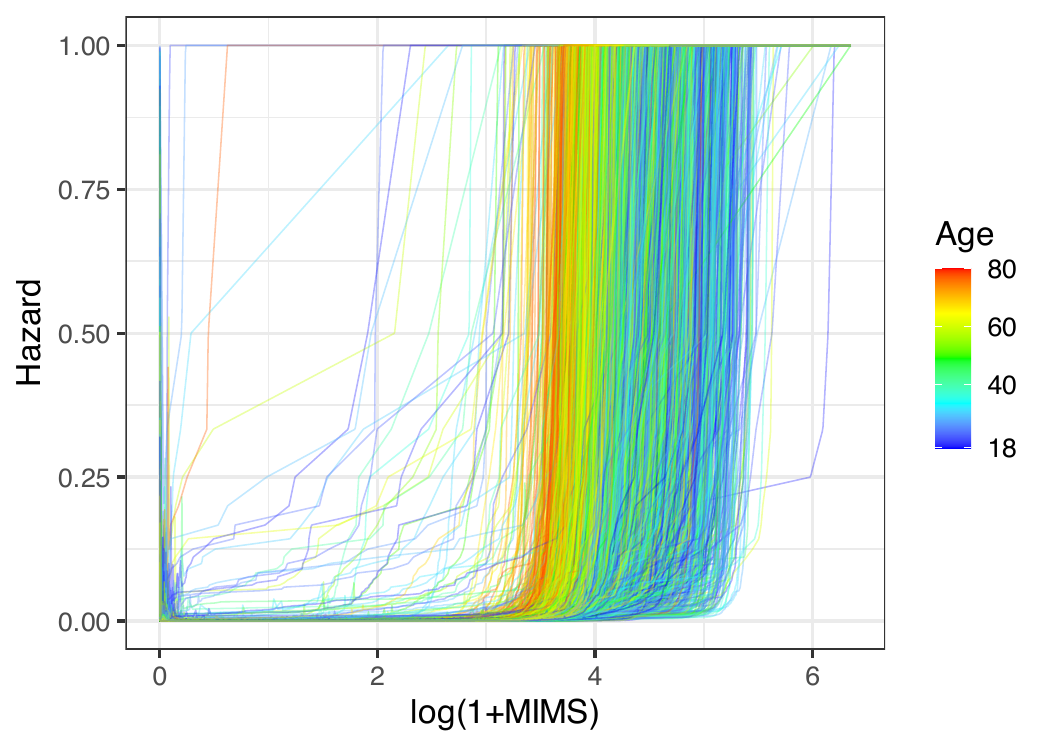}
        \caption{Subject-specific hazard curves indicate the progression over age.}
        \label{fig:all-logmims}
    \end{subfigure}
    
    \vspace{0.4cm}
    \begin{subfigure}[t]{0.48\textwidth}
        \centering
        \includegraphics[width=\textwidth]{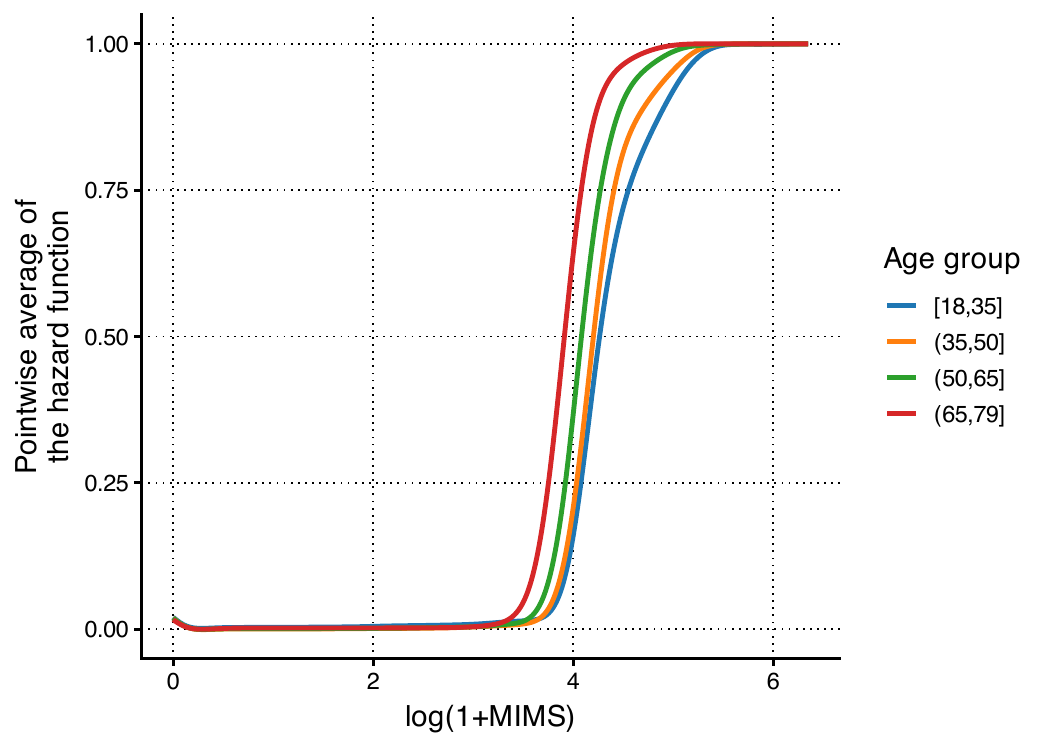}
        \caption{Group-wise mean hazard curves stratified by age groups.}
        \label{fig:age-logmims}
    \end{subfigure}
    \hfil
    \begin{subfigure}[t]{0.48\textwidth}
        \centering
        \includegraphics[width=\textwidth]{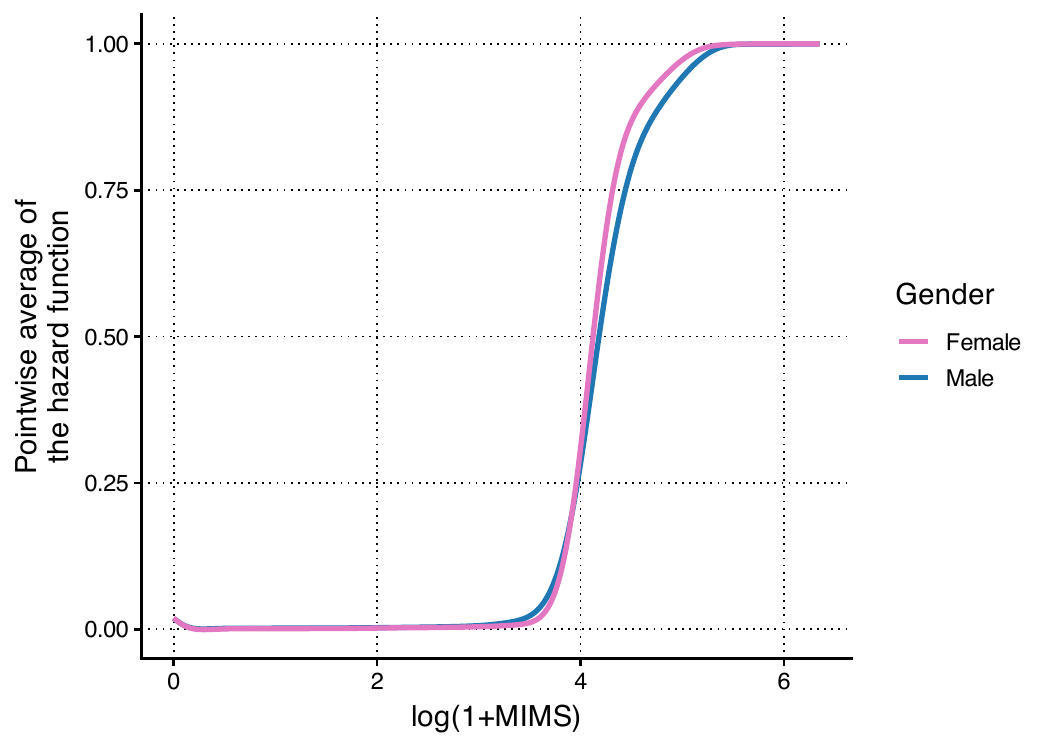}
        \caption{Group-wise mean hazard curves stratified by gender.}
        \label{fig:gender-logmims}
    \end{subfigure}
    
    \begin{subfigure}[t]{0.48\textwidth}
        \centering
        \includegraphics[width=\textwidth]{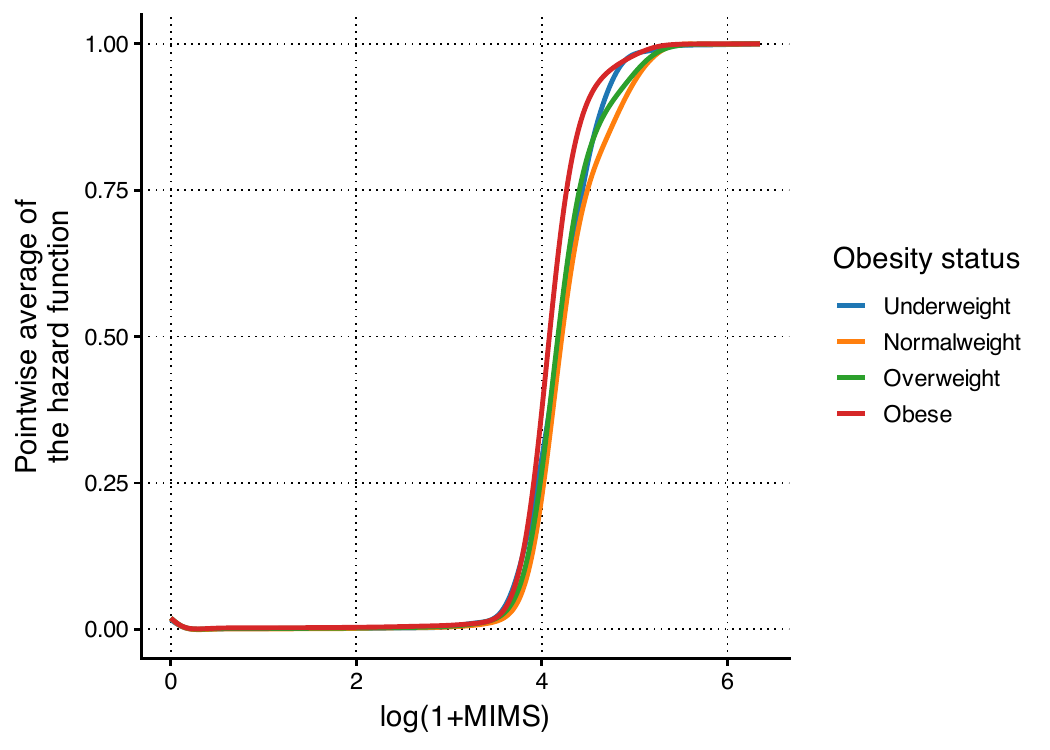}
        \caption{Group-wise mean hazard curves stratified by obesity status.}
        \label{fig:bmi-logmims}
    \end{subfigure}
    \hfill
    \begin{subfigure}[t]{0.48\textwidth}
        \centering
        \includegraphics[width=\textwidth]{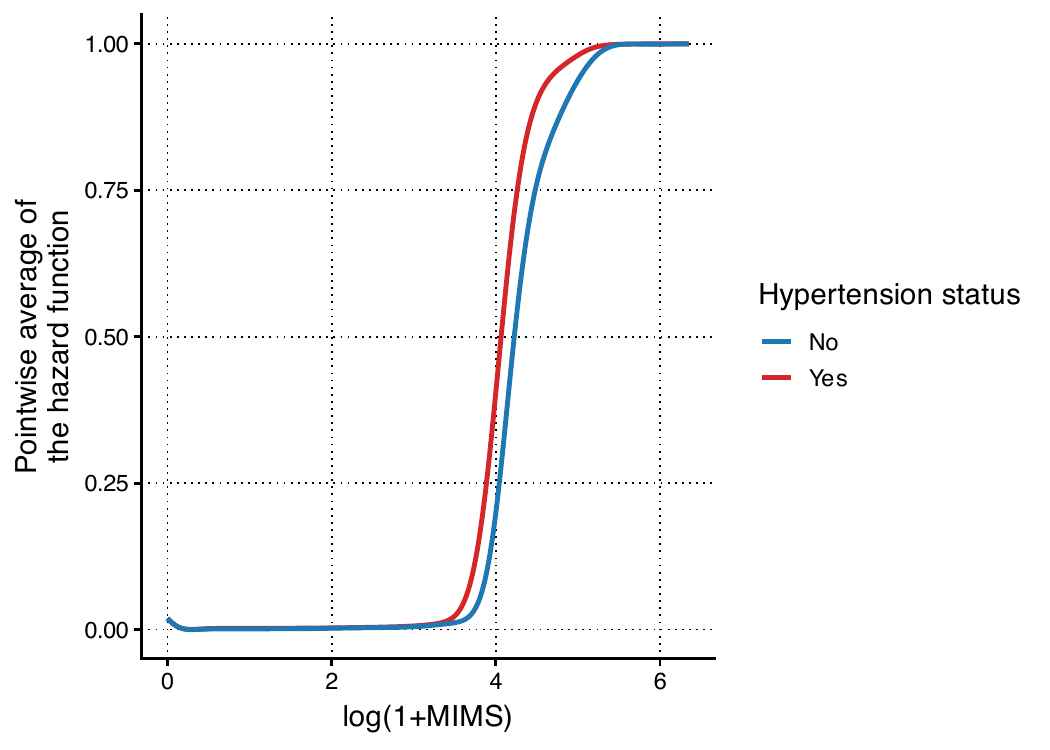}
        \caption{Group-wise mean hazard curves stratified by hypertension status.}
        \label{fig:htn-logmims}
    \end{subfigure}
    
    \caption{Observed subject-specific hazard functions estimated from log(1 + MIMS) activity intensities and their group-wise barycenters, illustrating differences in activity-intensity distributional shapes across subgroups.}

    \label{fig:descriptive_logmims}
\end{figure}

Figure \ref{fig:descriptive_logmims} displays the overall and group-wise graphical representations of the subject-specific hazard functions estimated from the log-transformed MIMS activity measures. An analogous set of graphical representations based on the untransformed MIMS activity measures is presented in Figure \ref{fig:descriptive_mims} in the Supplementary Material. The overall subject-specific hazard functions in Figures \ref{fig:all-logmims} and \ref{fig:all-mims} both exhibit a bathtub-shaped pattern; however, the width of the low-hazard region differs substantially between the log-transformed and original MIMS scales. On the log scale, the hazard function displays a longer flat region, indicating a broader and more stable range of moderate activity intensities. In contrast, the hazard function based on the original MIMS scale shows a shorter base, suggesting that activity intensities are more concentrated within a narrower band. The bathtub shape reflects a high concentration of observations at very low or sedentary activity levels, a relatively stable middle region, and an increasing hazard at high intensities due to the scarcity of extreme activity values. The longer flat region observed after log transformation implies that the survival function decays more gradually across moderate intensities, indicating greater heterogeneity in activity patterns. This suggests more sustained and variable engagement in light-to-moderate activity levels. By contrast, the original MIMS scale emphasizes extremely high-intensity values due to right skewness, which compresses the moderate range and leads to a shorter low-hazard region. The log transformation stabilizes variability, reduces skewness, and spreads moderate intensities more evenly across the scale. As a result, it provides a clearer representation of movement variability and central activity structure. Importantly, these structural differences vary across age groups, suggesting that the measurement scale may interact with demographic characteristics. Similar patterns are also observed across other clinical subgroups; however, we present the age-stratified results for illustration. These findings motivate a broader investigation of hazard-based distributional features across demographic and clinical subgroups.
\par
We computed point-wise group-specific mean hazard curves on a common grid. For each group $z \in \{1, \cdots, g\}$, the mean hazard function is defined as the empirical average of the subject-specific hazard functions within that group. This function summarizes the central tendency of the activity-intensity distribution at each intensity level and facilitates comparison of distributional shapes across groups. In Figures \ref{fig:age-logmims}–\ref{fig:htn-logmims} and \ref{fig:age-mims}–\ref{fig:htn-mims} in the Supplementary Material, we present graphical comparisons that demonstrate clear separation across demographic and clinical subgroups, facilitating the identification and interpretation of structural differences in activity-intensity distributions.
\par
The estimation of the model component functions $\mu(x \mid Z = z)$, $\phi_{k}(x \mid Z = z)$, and the subject-specific conditional hazard function in model \eqref{eq:haz} is carried out using the Fast Covariance Estimation (FACE) algorithm \citep{xiao2013fast, xiao2016fast}. Based on the subject-specific estimated hazard function $\widehat{h}_i(x| Z = z)$, $i=1,\ldots,n$ for group $z \in \{1, \cdots, g\}$, we evaluate it on a common grid $x_1,\ldots,x_L$ and we treat the collection $\{\widehat{h}_{i}(x_\ell| Z  = z): \ell = 1, \cdots, L\}$ as a densly observed functional data. Let $\bh_{iz} = \left( \widehat{h}_i(x_1| Z = z), \ldots, \widehat{h}_i(x_L| Z = z) \right)^\top$ denote the discretized hazard curve for subject $i$. 
A naive approach would compute the empirical covariance matrix $\widehat{\Sigma}_{z}$ based on $\bh_{iz}$ for each group. However, when the grid size $L$ is large, this procedure can be computationally intensive and numerically unstable due to noise in the estimated hazard functions. The FACE method addresses this issue by approximating the covariance surface through a reduced-rank smoother, ensuring computational efficiency, positive semi-definiteness of the estimated covariance matrix, and smooth eigenfunction estimates. Let $B \in \bbR^{L \times K}$ denote a B-spline basis matrix 
evaluated on $\{x_\ell\}_{\ell=1}^L$, where $K \ll L$. This basis provides a low-dimensional smooth representation 
of functions defined on the grid. For each group $\widehat{\Sigma}_{\mathrm{FACE}, z} = B \widehat{G}_{z} B^\top$ where $\widehat{G}_{z} \in \bbR^{K \times K}$ is obtained by minimizing a penalized least squares criterion
that balances fidelity to $\widehat{\Sigma}_{z}$ with smoothness, 
typically through a roughness penalty on spline coefficients. 
The smoothing parameter may be selected via generalized cross-validation 
or restricted maximum likelihood. Thus, perform the eigen-decomposition of $\widehat{G}_{z} = U_{z} \Lambda_{z} U_{z}^\top$ so that we want to normalize the eigen-functions on the original grid are then given by $\widehat{\phi}_k(x| Z = z) = \sum\limits_{\ell=1}^{K} U_{\ell kz} B_\ell(x)$, with corresponding eigenvalues $\widehat{\lambda}_{kz}$, where $U_{\ell kz}$ is the $(\ell, k)$-th entry of $U_{z}$.
For each subject $i = 1, \cdots, n$, and group $z \in \{1, \cdots, g\}$, compute the scores $\widehat{\xi}_{ikz} = 
\sum\limits_{\ell=1}^{L} 
\left( \widehat{h}_i(x_\ell| Z = z) - \bar{h}(x_\ell| Z = z) \right)
\widehat{\phi}_k(x_\ell) \Delta x_{\ell}$, where $\bar{h}(x_\ell| Z = z)$ is the groups specific point-wise averages. This provides a low-dimensional representation of the hazard curve. The resulting eigenfunctions $\widehat{\phi}_k(x| Z = z)$ 
describes the dominant modes of variation in the activity-intensity hazard functions, 
while the scores $\widehat{\xi}_{ik}$ 
summarize subject-specific deviations from the mean structure. Let $\widehat{\lambda}_{1z} \ge \widehat{\lambda}_{2z} \ge \cdots \ge \widehat{\lambda}_{rz}$ 
denote the estimated eigenvalues of the smoothed covariance operator. 
Define the fraction of variance explained (FVE) by the first $K$ components as $\mathrm{FVE}(K)
=
\frac{\sum\limits_{k=1}^{K}\widehat{\lambda}_k}
{\sum\limits_{k=1}^{r}\widehat{\lambda}_k}$. We want to retain the 95\% of the total variability; therefore, we select the number of retained components as $\widehat{K} =
\min \left\{
K \in \{1,\ldots,r\} :
\frac{\sum\limits_{k=1}^{K}\widehat{\lambda}_k}
{\sum\limits_{k=1}^{r}\widehat{\lambda}_k}
\ge \eta
\right\}$. This criterion provides an effective balance between dimensionality reduction and retention of the dominant patterns in the data. 

\section{Results}
\label{sec:results}
In this section, we present a detailed analysis of the NHANES data, examining the distributional characteristics of physical activity intensity using the hazard-function representation introduced in Section \ref{sec:model} and derived from accelerometry-based MIMS measures. We first estimate subject-specific hazard functions to capture the shape of the activity-intensity distribution for each individual. Treating these hazard functions as functional observations, we summarize the overall patterns and examine differences across demographic and clinical subgroups. In particular, we compare group-level mean hazard curves and their barycenters to identify structural differences in activity-intensity distributions. These graphical and functional summaries provide insight into heterogeneity in activity patterns that conventional summary measures of physical activity may not capture. 
\par
To gain intuition about the behavior of the hazard functions, we first examine the patterns shown in Figures \ref{fig:descriptive_logmims} and \ref{fig:descriptive_mims} in the Supplementary Material. The overall hazard function is relatively low at lower activity-intensity levels and increases as the intensity rises. This pattern reflects the fact that individuals spend a larger proportion of their day in sedentary and light-intensity physical activity (LIPA), leading to greater concentration of observations in these ranges. In contrast, the hazard increases at higher intensity levels because individuals spend relatively little time engaging in high-intensity activity. At lower intensity levels, the mean hazard curves are nearly identical across the different age categories, indicating similar distributional patterns of sedentary and light-intensity physical activity. However, at higher intensity levels, the older age groups exhibit higher hazard rates than the younger groups. This pattern suggests that older individuals spend relatively less time engaging in high-intensity physical activities, whereas younger individuals tend to remain in higher-intensity activity levels for longer periods.
The hazard curves for males and females closely overlap at lower intensity levels, indicating a similar distribution of activity intensity within the sedentary and light-intensity physical activity (LIPA) ranges for both groups. However, as the intensity level increases, the female group exhibits slightly higher hazard rates than the male group. This suggests that women spend relatively less time engaging in high-intensity physical activities compared with men. At lower activity-intensity levels, the mean hazard curves overlap closely across all BMI groups, indicating similar distributions of sedentary and light-intensity physical activity. Differences between the groups become more pronounced at higher intensity levels. In particular, the obese group exhibits the highest hazard rates, followed by the underweight and overweight groups, while the normal-weight group consistently shows the lowest hazard values. The curves for the underweight and overweight groups also exhibit partial overlap. These patterns suggest that obese individuals spend relatively fewer minutes engaging in high-intensity physical activity, whereas individuals in the normal-weight category tend to spend more time at these higher intensity levels. The hazard curves for hypertensive and non-hypertensive individuals closely overlap at lower activity-intensity levels, indicating similar distributions of sedentary and light-intensity physical activity across the two groups. In contrast, the separation between the curves becomes more pronounced at higher intensity levels. In this region, hypertensive individuals exhibit higher hazard rates than non-hypertensive individuals, suggesting that hypertensive individuals spend relatively less time engaging in high-intensity physical activity.
\par
Now we have performed FPCA (see, model \eqref{eq:fpca}) based on the entire data (without consideration of the demographic and clinical groups) to capture the dominant modes of variation in the subject-specific hazard functions and to provide a parsimonious representation of the functional data. Since each subject’s hazard curve is evaluated on a dense grid, the resulting functions are inherently high-dimensional and may exhibit substantial correlation across the activity-intensity domain. FPCA addresses this issue by decomposing the covariance structure of the hazard functions into orthogonal eigenfunctions and associated scores, allowing the complex functional variability to be summarized by a small number of principal components. This dimension reduction facilitates interpretation of the main patterns of variation in activity-intensity distributions and enables efficient comparison across demographic and clinical subgroups. 
\par
\begin{figure}[htbp]
  \centering
  \begin{subfigure}[t]{0.49\textwidth}
    \centering
    \includegraphics[width=\textwidth]{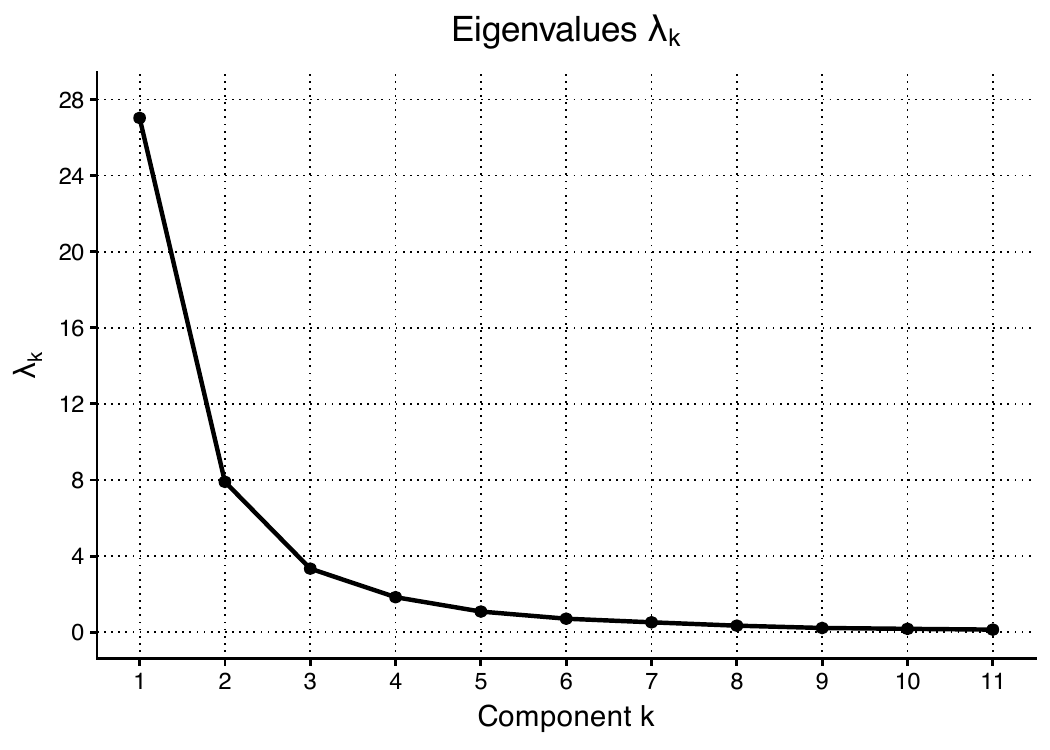}
    \caption{Model based on MIMS}
    \label{fig:fpca_eigenvalues_mims}
  \end{subfigure}
  \hfill
  \begin{subfigure}[t]{0.49\textwidth}
    \centering
    \includegraphics[width=\textwidth]{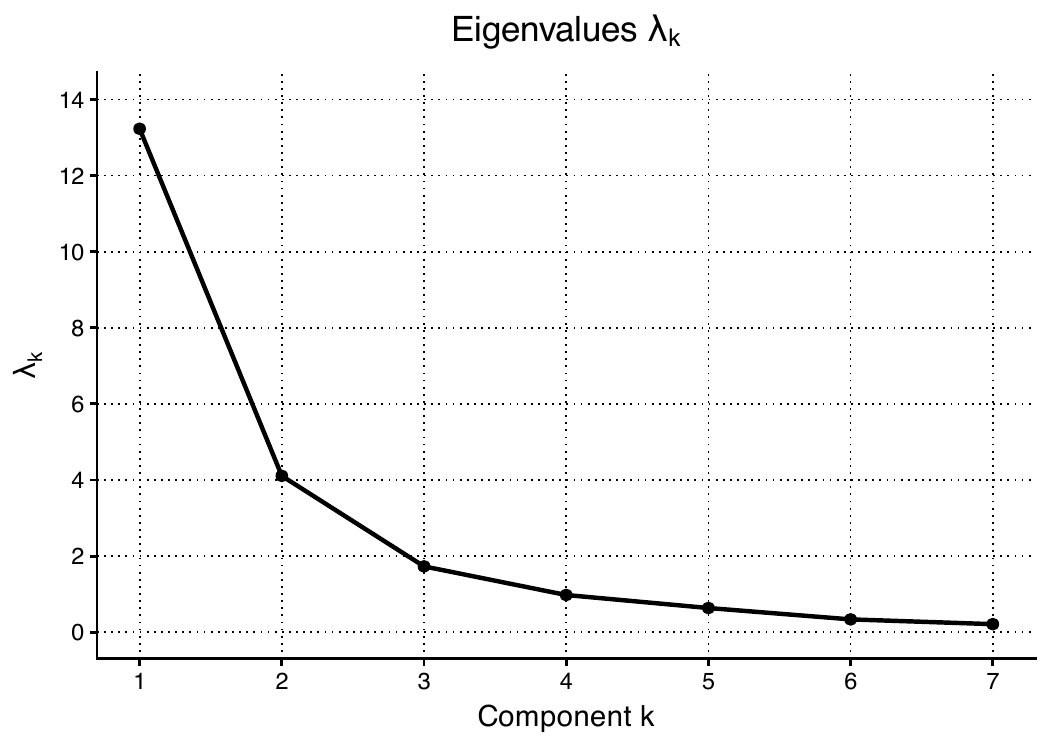}
    \caption{Model based on log(1+MIMS)}
    \label{fig:fpca_eigenvalues_logmims}
  \end{subfigure}
  \caption{Scree plots for selecting the number of principal components in FPCA for hazard functions based on MIMS on the raw scale (left) and the log-transformed intensity scale (right).}
  \label{fig:fpca_eigenvalues_both}
\end{figure}

Group-wise FPCA (see, model \eqref{eq:haz}) is performed to investigate whether the dominant modes of variation in the hazard functions differ across demographic and clinical subgroups. While standard FPCA summarizes the overall variability in the entire sample, it may mask subgroup-specific structural patterns when the functional distributions differ substantially between groups. By conducting FPCA separately within each group, we allow the covariance structure and the associated eigenfunctions to adapt to group-specific activity-intensity distributions. This enables us to identify distinct patterns of variability and heterogeneity within each subgroup and provides a more interpretable characterization of how physical activity intensity distributions differ across demographic and clinical populations. Here, we have fitted two seperate model, one with MIMS in the original scale and log-transformed MIMS to address the skewness in the intensity and stabilize across the intensity range. It is important to mention that based on Figure \ref{fig:descriptive_logmims} and \ref{fig:descriptive_mims} in the Supplementary Material, we have observed that the raw MIMS scale tends to emphasize extremely high-intensity values, which can compress the moderate activity region. In contrast, the log-transformed scale spreads moderate intensities, making it easier to identify structural features of the activity distribution, such as the flat middle region of the bathtub-shaped hazard. Comparing the two scales also provides a sensitivity analysis to assess whether the observed distributional patterns depend on the measurement scale. For ease of interpretation, we fix the number of knots at 20. The number of eigen-components is then determined using the fraction of variance explained (FVE) criterion. Specifically, we set the cumulative variance explained threshold at 0.95 and select the smallest number of eigenvalues required to reach this level. Figure \ref{fig:fpca_eigenvalues_both} presents the scree plot of the estimated eigenvalues, illustrating the fraction of variance explained by each principal component and guiding the selection of the number of retained components based on the FVE criterion. FPCA applied to the hazard functions on both the raw MIMS scale and the transformed scale log(1+MIMS) indicates that the functional variability is largely concentrated in the leading components. On the raw MIMS scale, the first eigenvalue $(\lambda_{1} \approx 11.76)$ explains approximately 66\% of the total functional variance, indicating that a dominant mode of variation governs the shape of the hazard curves across subjects. The second and third eigenvalues account for an additional 19.2\% and 7.4\% of the variability, respectively, so that the first three principal components explain approximately 92.6\% of the total variation. Under the pre-specified FVE threshold of 0.95, four components are sufficient to capture the majority of the functional variability. A similar pattern is observed on the log(1+MIMS) scale, although the transformation redistributes variability more evenly across the leading components due to the reduction of skewness in the activity-intensity distribution. Consequently, the principal component representation provides a low-dimensional yet accurate approximation of the hazard-function variability on both scales. 
Table \ref{tab:strat_eigs} presents the leading eigenvalues obtained from stratified FPCA together with the corresponding FVE for the first three components under the raw and log-transformed MIMS hazard representations. The percentages reported in parentheses denote the proportion of within-stratum functional variability captured by each component, allowing comparison of the dominant modes of variation across demographic and clinical subgroups. Across both activity-intensity scales and all strata, the first principal component explains the largest proportion of functional variability, accounting for approximately 60–70\% of the total variance. The second and third principal components explain smaller but still non-negligible proportions of the variance (see, Table~\ref{tab:strat_eigs}). The estimated eigenfunctions exhibit broadly similar structural patterns across subgroups within comparable intensity regions; however, they differ in the magnitude and localization of these features. Since FPCA eigenfunctions are identifiable only up to a sign change, comparisons across strata focus on the locations of peaks, troughs, and dominant intensity regions rather than the direction of the curves. These differences indicate meaningful heterogeneity in the distributional structure of physical activity intensity across demographic and clinical subgroups.

\begin{table}[htbp]
\centering
\caption{Stratified FPCA eigenvalues and fraction of variance explained (FVE) for the first three components under hazard representations based on raw MIMS and log-transformed MIMS. Percentages in parentheses denote within-stratum FVE.}
\label{tab:strat_eigs}

\begin{threeparttable}
\begin{adjustbox}{max width=\textwidth,center}
\begin{tabular}{llccc ccc}
\toprule
\multirow{2}{*}{Covariate} & \multirow{2}{*}{Category} &
\multicolumn{3}{c}{Raw MIMS} & \multicolumn{3}{c}{Log-transformed MIMS} \\
\cmidrule(lr){3-5}\cmidrule(lr){6-8}
& & $\lambda_1$ (\%) & $\lambda_2$ (\%) & $\lambda_3$ (\%) &
$\lambda_1$ (\%) & $\lambda_2$ (\%) & $\lambda_3$ (\%) \\
\midrule

\multirow{2}{*}{Gender}
& Male   & 31.27 (62.98) & 8.96 (18.06) & 3.72 (7.49) & 14.83 (61.94) & 4.56 (19.05) & 1.97 (8.24) \\
& Female & 22.38 (61.89) & 6.90 (19.08) & 2.82 (7.81) & 11.43 (62.81) & 3.65 (20.06) & 1.45 (7.95) \\
\midrule

\multirow{4}{*}{Age}
& [18--35] & 35.21 (67.48) & 7.86 (15.07) & 3.48 (6.68) & 14.81 (64.54) & 3.62 (15.79) & 2.12 (9.23) \\
& (35--50] & 26.09 (64.33) & 7.08 (17.45) & 3.02 (7.44) & 11.84 (65.77) & 3.47 (19.26) & 1.31 (7.29) \\
& (50--65] & 18.03 (60.58) & 5.69 (19.11) & 2.54 (8.53) & 10.19 (63.80) & 3.00 (18.81) & 1.39 (8.73) \\
& (65--79] & 13.11 (60.65) & 4.42 (20.43) & 1.94 (8.98) & 9.33 (62.17) & 2.97 (19.83) & 1.22 (8.12) \\
\midrule

\multirow{4}{*}{BMI}
& Underweight   & 23.95 (61.77) & 7.10 (18.31) & 3.27 (8.44) & 13.51 (66.10) & 3.72 (18.21) & 1.50 (7.34) \\
& Normal weight & 33.94 (66.64) & 8.09 (15.88) & 3.72 (7.30) & 15.22 (66.83) & 4.00 (17.55) & 1.60 (7.05) \\
& Overweight    & 28.37 (62.17) & 8.30 (18.19) & 3.50 (7.66) & 13.46 (62.17) & 4.25 (19.64) & 1.77 (8.17) \\
& Obese         & 18.53 (58.59) & 6.45 (20.38) & 2.85 (9.01) & 10.54 (58.94) & 3.50 (19.60) & 1.74 (9.74) \\
\midrule

\multirow{2}{*}{Hypertension}
& No  & 31.70 (66.06) & 7.77 (16.20) & 3.43 (7.16) & 14.00 (65.99) & 3.69 (17.41) & 1.56 (7.35) \\
& Yes & 19.53 (58.72) & 6.82 (20.49) & 2.88 (8.65) & 11.24 (60.02) & 3.68 (19.63) & 1.73 (9.23) \\
\bottomrule
\end{tabular}
\end{adjustbox}

\begin{tablenotes}\footnotesize
\item[] 
\parbox{0.97\textwidth}{%
\hangindent=2.6em
\hangafter=1
\textit{Note:} Eigenfunctions are identifiable up to sign; sign reversals across strata do not imply different modes.
}
\end{tablenotes}
\end{threeparttable}
\end{table}

\begin{figure}[htbp]
    \centering
    \begin{subfigure}[t]{\textwidth}
        \centering
        \includegraphics[width=0.95\textwidth]{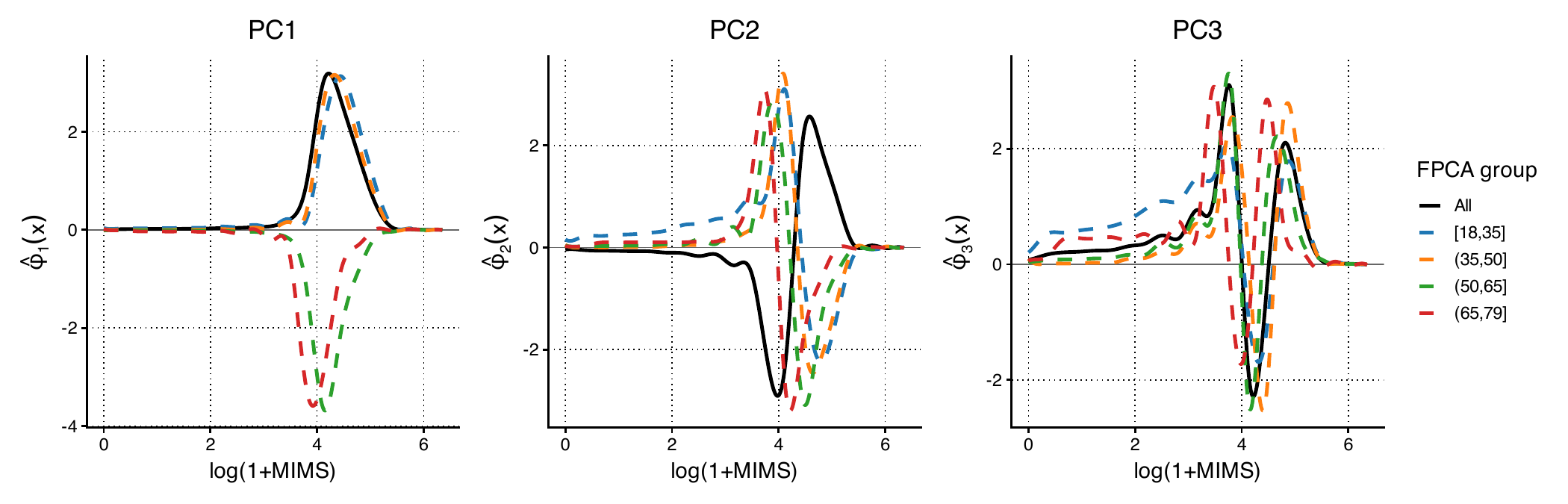}
        \caption{FPCA eigenfunctions stratified by age group}
    \end{subfigure}
    
    \begin{subfigure}[t]{\textwidth}
        \centering
        \includegraphics[width=\textwidth]{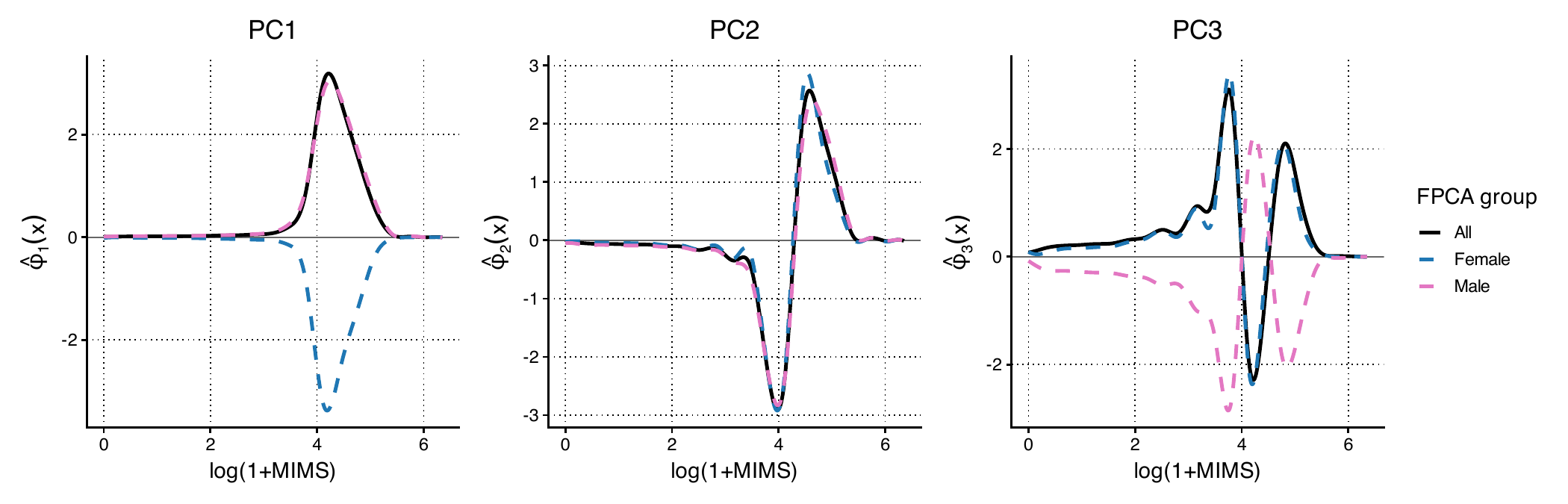}
        \caption{FPCA eigenfunctions stratified by gender}
    \end{subfigure}
    
    \begin{subfigure}[t]{\textwidth}
        \centering
           \includegraphics[width=0.95\textwidth]{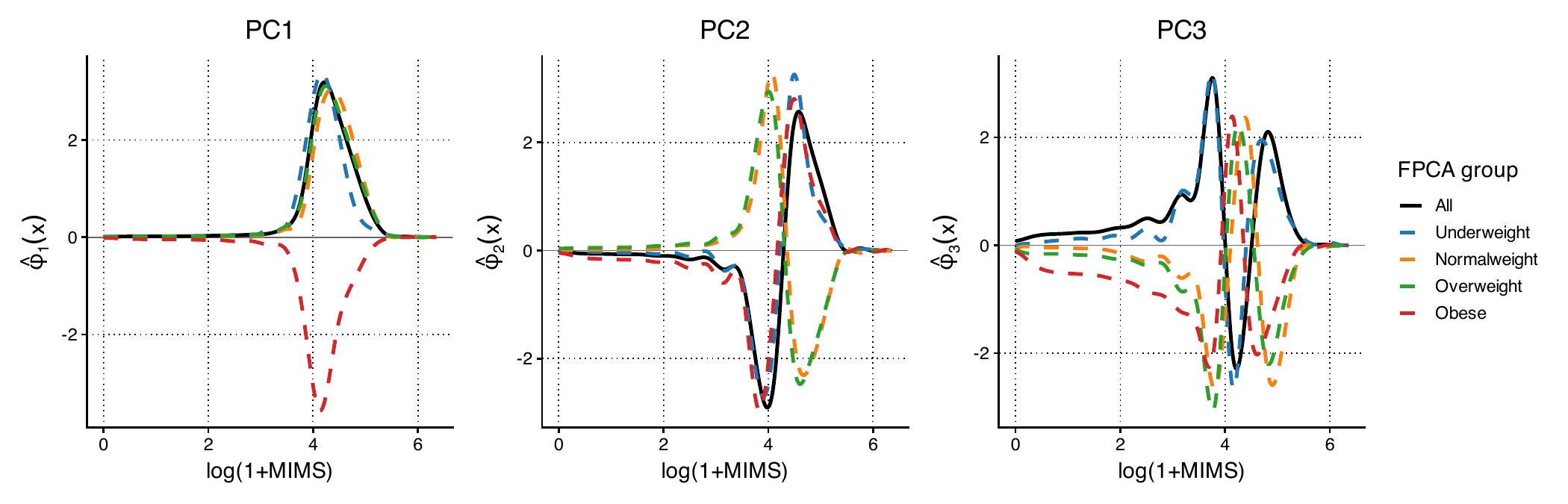}
        \caption{FPCA eigenfunctions stratified by obesity status}

    \end{subfigure}
    \begin{subfigure}[t]{\textwidth}
        \centering
        \includegraphics[width=0.95\textwidth]{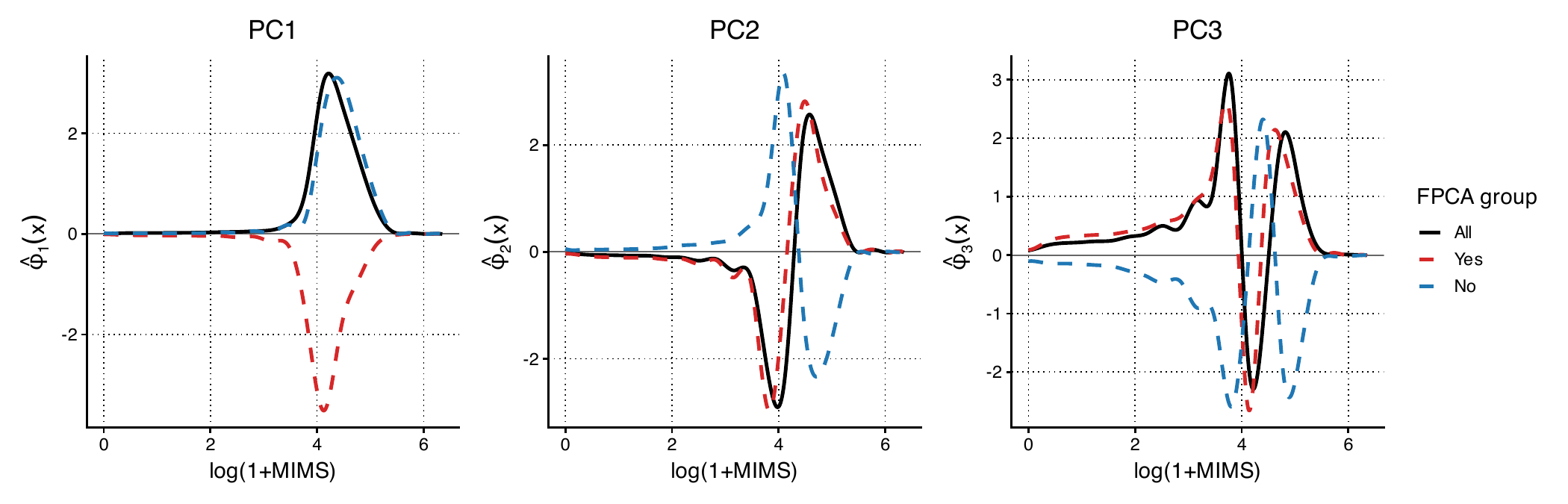}
        \caption{FPCA eigenfunctions stratified by hypertension status}
    \end{subfigure}
    
    \caption{First three eigen-functions using models \eqref{eq:fpca} and \eqref{eq:haz} based on log-transformed MIMS.}

    \label{fig:fpca_eigenfunctions_logmims}
\end{figure}

The first three eigenfunctions estimated from FPCA of the activity–intensity hazard functions evaluated on the log-transformed MIMS scale and raw MIMS are displayed in Figures \ref{fig:fpca_eigenfunctions_logmims} and \ref{fig:fpca_eigenfunctions_mims} in the Supplementary Material. Since eigenfunctions are identifiable only up to a sign, we impose the constraint $\int \widehat{\phi}_{k}(x) \widehat{\phi}_{k}(x |Z = z) dx \geq 0$ to ensure consistent orientation and facilitate meaningful comparisons across demographic and clinical groups. Both representations—based on the raw MIMS scale and the log-transformed intensity scale— describe the same underlying activity–intensity hazard behavior. Nevertheless, the rescaling of intensity redistributes functional variability across the domain, leading to differences in the estimated eigenfunctions. 
\par
The first eigenfunction (PC1) estimated based on the log-transformed MIMS shows its strongest magnitude at moderate-to-high activity intensities. This suggests that individuals differ primarily in how the hazard increases with increasing activity intensity. Participants with higher PC1 scores tend to exhibit an earlier and sharper increase in the hazard at higher intensity levels, whereas lower PC1 scores indicate relatively less time spent in higher-intensity physical activity. Overall, the first principal component captures the dominant structural variation in the activity-intensity hazard curves, reflecting how individuals differ in the distribution of time spent across intensity levels (See, Figure \ref{fig:fpca_eigenfunctions_logmims}).
In contrast, the first eigenfunction (PC1) estimated from the MIMS scale exhibits its largest magnitude at low to moderate intensity levels, followed by a rapid decay as intensity increases. This pattern indicates that the primary source of between-individual variability occurs in the lower intensity range, where individuals spend the majority of their daily time. Individuals with higher PC1 scores tend to exhibit greater variability in the low-to-moderate intensity range, whereas individuals with lower PC1 scores show more stable activity patterns within these intensity levels (See, Figure \ref{fig:fpca_eigenfunctions_mims}). 

\par
The second eigenfunction (PC2) on the MIMS scale primarily reflects variation in the low-to-moderate intensity region. This component captures differences in how individuals transition from very low activity intensities toward higher intensity levels. Individuals with positive PC2 scores tend to exhibit more gradual changes in the hazard function within the low-intensity region, whereas negative PC2 scores correspond to sharper changes in hazard at these intensity levels. On the log-transformed MIMS scale, the second eigenfunction contrasts the moderate and higher intensity regions, as indicated by opposite signs across these ranges. This component characterizes heterogeneity in how the hazard increases across activity intensities. Individuals with positive PC2 scores tend to exhibit a more gradual increase in hazard across the intensity domain, whereas negative PC2 scores correspond to delayed but sharper increases in hazard at higher intensity levels. Thus, the second principal component captures differences in how rapidly individuals transition across activity-intensity levels, distinguishing gradual versus abrupt increases in the hazard function.
\par
The third eigenfunction (PC3) represents participant-specific variation beyond the dominant patterns captured by the first two principal components. This component reflects more localized deviations in the hazard functions, primarily concentrated at higher activity intensities, and accounts for a relatively smaller proportion of the total variance, particularly on the log-transformed scale.
\par
Overall, these eigenfunctions characterize the principal modes of variability in the activity–intensity hazard distributions. The variability is driven mainly by structured differences in the moderate-to-vigorous intensity range, whereas the hazard functions at lower intensity levels remain relatively similar across individuals.
\par
Age-stratified eigenfunctions show clearer variation than those stratified by gender. On the raw scale, PC1 remains dominant in the low-to-moderate intensity region; however, the shape and localization of the main peak differ across age groups, suggesting age-dependent differences in how variability is distributed within lower intensity levels (Figure~\ref{fig:fpca_eigenfunctions_mims}, panel~(a)). On the log-transformed scale, all age groups exhibit strong variation centered in a common mid-to-higher log-intensity region. Differences across age groups are primarily observed in the alignment and localization of higher-order modes (Figure~\ref{fig:fpca_eigenfunctions_logmims}, panel~(a)).
The gender-specific eigenfunctions broadly consistent with the overall eigenfunctions on the raw MIMS hazard scale. However, the male PC1 shows a sign flip at lower intensity levels, indicating some gender-specific differences in the primary mode of variation. The leading eigenfunction (PC1) concentrates variability in the low-to-moderate intensity region, with near-zero loadings at higher intensities. This suggests that between-participant variability is primarily driven by how activity mass is distributed within the lower intensity range (Figure~\ref{fig:fpca_eigenfunctions_mims}, panel~(b)). On the log(1 + MIMS) hazard scale, the eigenfunctions again exhibit similar dominant structures, with variability centered around the mid-to-higher log-intensity region. Apparent sign differences across subgroups arise from the inherent sign indeterminacy of FPCA (Figure~\ref{fig:fpca_eigenfunctions_logmims}, panel~(b)). Across BMI categories, the first principal component shows variability primarily in the low-to-moderate intensity region on the raw MIMS scale, while the second and third principal components reveal clearer separation across BMI groups through localized differences at adjacent intensity levels (Figure~\ref{fig:fpca_eigenfunctions_mims}, panel~(c)). On the log(1+MIMS) scale, variability is concentrated in the mid-to-higher intensity region, with BMI-related differences arising mainly from where within this region the variability is localized (Figure~\ref{fig:fpca_eigenfunctions_logmims}, panel~(c)). Hypertension-stratified eigenfunctions are broadly aligned for PC1–PC2 on both scales, indicating similar dominant modes of variability, though PC2 exhibits contrasting patterns between hypertensive and non-hypertensive participants. Differences are more pronounced in the higher-order components, where hypertensive and non-hypertensive participants exhibit distinct localized contrast patterns across low-to-moderate intensity levels on the raw scale and differing alignment in the moderate-to-high log-intensity region on the log-transformed scale (Figure~\ref{fig:fpca_eigenfunctions_logmims}, panel~(d)). These results suggest that hypertension status is reflected less in the primary mode of variation and more in higher-order redistribution patterns across activity-intensity bands.
Overall, the FPCA results reveal that variability in activity–intensity hazard functions is primarily driven by structured differences in the low-to-moderate and moderate-to-high intensity regions of physical activity. Across both the raw MIMS and log-transformed intensity scales, the first few principal components capture the dominant modes of variation, indicating that most between-participant differences arise from how individuals distribute their activity across intensity levels. While the overall functional patterns are broadly consistent across demographic and clinical subgroups, stratified analyses show that the magnitude and localization of variability differ across groups such as age, BMI, and hypertension status. In particular, subgroup-specific differences are more evident in higher-order components, reflecting localized redistribution of activity intensity rather than changes in the primary mode of variation. These findings highlight that heterogeneity in physical activity behavior is largely governed by differences in engagement at moderate and higher intensity levels, while sedentary and light-intensity activity patterns remain relatively similar across individuals. From a methodological perspective, representing activity intensity through hazard functions and analyzing them via FPCA provides a flexible framework for capturing distributional characteristics of physical activity and uncovering structured variability that may not be evident from traditional summary measures of activity behavior.

\section{Discussion}
\label{sec:discussion}
In this paper, we analyzed wrist-worn accelerometry data to study physical activity intensity distributions and found their dominant patterns of variability using FPCA. We presented physical activity intensity in the form of hazard curves and found that the heterogeneity mainly occurs at moderate-to-high intensity activity levels rather than sedentary or light-intensity regions. Following this hazard-based distributional approach, we captured differences in physical activities across demographic and clinical subgroups, including age, sex, obesity status, and hypertension; traditional mean-based approaches often neglect these differences.

Our findings clearly highlighted that the proposed hazard-based functional representation characterizes the complexity of physical activity behavior compared with conventional scalar summaries such as total activity counts, average activity, or total moderate-to-vigorous physical activity duration. Conventional scalar measures often overlook important information about the distributional structure of activity intensity across the day. Individuals with similar average activity may still exhibit different distributions of activity intensity. The results show that the differences between subgroups are more evident at moderate and higher intensities of activities. The estimated hazard functions showed clear separation between demographic and clinical populations. These findings are consistent with previous studies highlighting the value of distribution-based representations of accelerometry data beyond traditional average-based measures \citep{ghosal2023distributional, niyogi2024hazard, niyogi2026quantifying}. Our findings also support previous work emphasizing distribution-based modeling approaches for better characterizing physical activity behavior from accelerometry data \citep{ghosal2023distributional, mendez2025functional, niyogi2024hazard}.

From a physiological perspective, the subgroup differences align with established behavioral and clinical patterns associated with physical activity. Older individuals showed higher hazards at higher intensity levels, indicating reduced engagement in vigorous physical activity. This finding is consistent with existing literature suggesting that physical activity generally reduces with age \citep{hyde2013enhancing, mohd2021review}. Similarly, obese individuals demonstrated a higher hazard at higher intensity levels compared with the normal-weight individuals, reflecting reduced participation in higher-intensity activity. This pattern may be explained by fatigue, reduced physical endurance, and the biomechanical and metabolic burdens associated with obesity \citep{jakicic2011obesity, ortega2019cardiorespiratory, sagelv2021bidirectional}. Hypertensive participants also had higher hazards in higher intensity regions, suggesting lower participation in vigorous physical activity, which is consistent with previous findings linking physical inactivity with hypertension and cardiovascular risk \citep{rossi2012impact, hayes2022physical, vieira2024barriers}.

A very notable finding of this study is that dominant modes of variability were different between the raw MIMS and log-transformed MIMS scales. The first principal component on the original MIMS scale mainly captured variability within the low-to-moderate activity intensity range. In contrast, the log-transformed representation revealed the dominant functional variability toward the moderate-to-high intensity regions. This difference reflects the strong right-skewed nature of accelerometry intensity distributions. So, transformation in logarithmic scale reduced the variability across the intensity range and provided a clearer view of structural heterogeneity in physical activity. Similar distribution-based and functional representations have recently been explored in wearable sensor and accelerometry studies using quantile- and hazard-based functional approaches \citep{mendez2025functional}. Both scales (raw and log-transformed) appear to provide complementary information. The original scale captures the effects of extreme activity intensities, and the transformed scale improves interpretability in the moderate intensity range.

FPCA results showed that the majority of the functional variability was explained by the first few principal components, where the first component accounted for approximately 60--70\% of the total variability across the groups. This suggests that the major heterogeneity in physical activity behavior can be captured by a comparatively low-dimensional functional framework. Although the first principal component remained largely consistent across groups, the higher-order components highlighted localized variations specific to individual demographic and clinical subgroups. These higher-order components likely reveal more localized and subtle differences in how activity intensity is distributed across individuals. The results tell us that differences in physical activity behavior are not only a function of differences in overall activity level but also of how activity is distributed across intensity regions. This interpretation is consistent with previous FPCA and distribution-based studies showing that the leading functional components often capture dominant global variability patterns, whereas higher-order components reflect more localized structural variations \citep{chiou2009modeling}. Recent work on accelerometry-based distributional representations has also emphasized that hazard-based functional representations can capture clinically meaningful differences in activity behavior more effectively than conventional summary measures \citep{niyogi2024hazard}.

Overall, this study demonstrates that hazard-based functional representations provide a useful framework for characterizing physical activity behavior from wearable accelerometry data. The growing availability of high-resolution wearable sensor data presents new opportunities for studying human behavior at an unprecedented level of detail. Realizing the full potential of these data requires analytical approaches that move beyond traditional low-dimensional summaries and account for the complex distributional structure of physical activity. The proposed framework represents one step in this direction and provides a foundation for future methodological developments that integrate functional, distributional, and longitudinal aspects of wearable-derived activity measurements. Ultimately, these advances may contribute to more precise behavioral phenotyping and a deeper understanding of the relationships between physical activity patterns and health outcomes.

\section*{Supplementary Material}
\renewcommand{\thefigure}{S\arabic{figure}}
\renewcommand{\thetable}{S\arabic{table}}
\setcounter{figure}{0}
\setcounter{table}{0}

This supplementary document contains the additional figures and supporting materials.

\begin{figure}[htbp]
    \centering
    \begin{subfigure}[t]{0.48\textwidth}
        \centering
        \includegraphics[width=\textwidth]{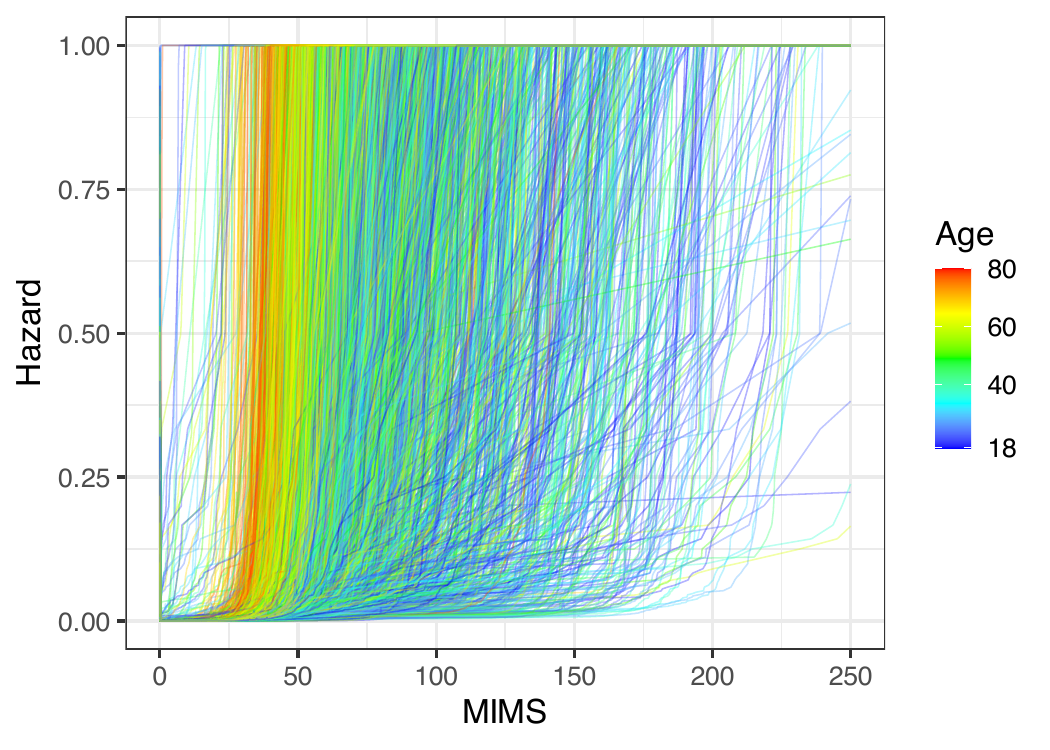}
        \caption{Subject-specific hazard curves indicate the progression over age.}
        \label{fig:all-mims}
    \end{subfigure}
    
    \vspace{0.4cm}
    \begin{subfigure}[t]{0.48\textwidth}
        \centering
        \includegraphics[width=\textwidth]{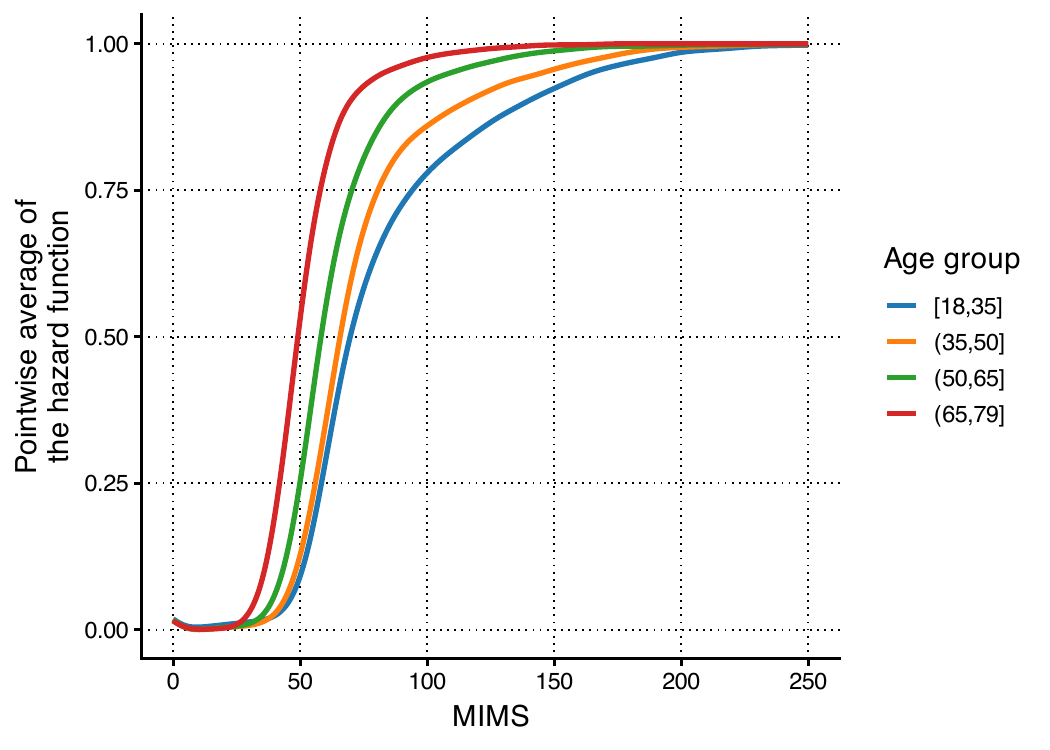}
        \caption{Group-wise mean hazard curves stratified by age groups.}
        \label{fig:age-mims}
    \end{subfigure}
    \hfil
    \begin{subfigure}[t]{0.48\textwidth}
        \centering
        \includegraphics[width=\textwidth]{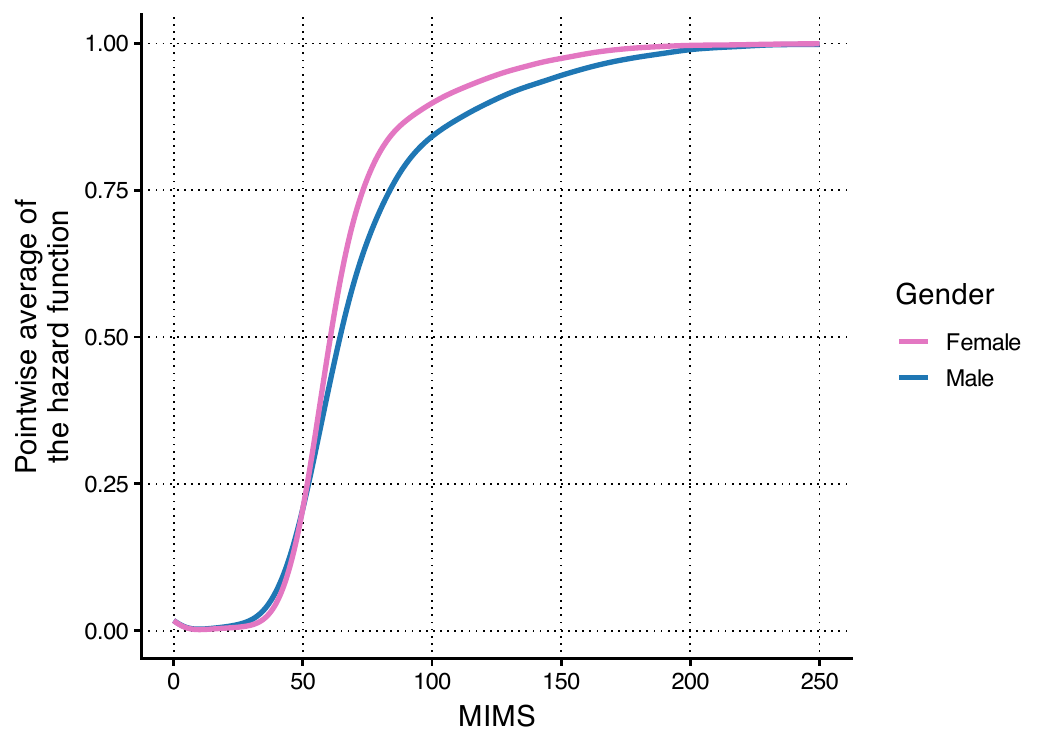}
        \caption{Group-wise mean hazard curves stratified by gender.}
        \label{fig:gender-mims}
    \end{subfigure}
    
    \begin{subfigure}[t]{0.48\textwidth}
        \centering
        \includegraphics[width=\textwidth]{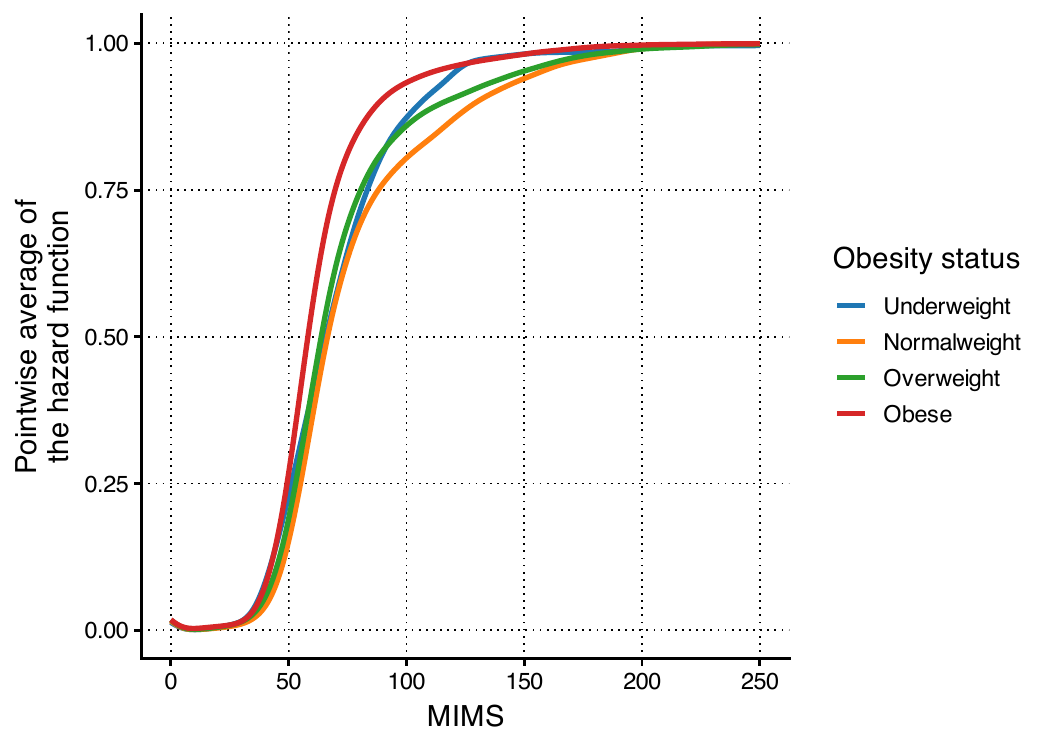}
        \caption{Group-wise mean hazard curves stratified by obesity status.}
        \label{fig:bmi-mims}
    \end{subfigure}
    \hfill
    \begin{subfigure}[t]{0.48\textwidth}
        \centering
        \includegraphics[width=\textwidth]{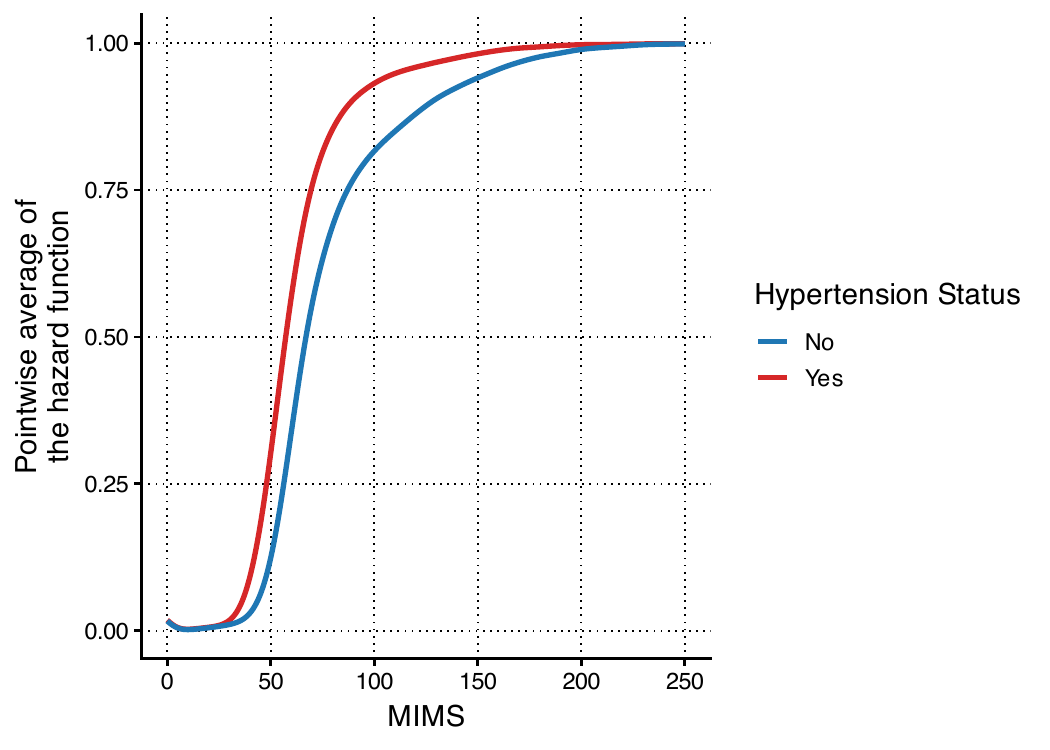}
        \caption{Group-wise mean hazard curves stratified by hypertension status.}
        \label{fig:htn-mims}
    \end{subfigure}
    
    \caption{Observed subject-specific hazard functions estimated from MIMS activity intensities and their group-wise barycenters, illustrating differences in activity-intensity distributional shapes across subgroups.}

    \label{fig:descriptive_mims}
\end{figure}

\begin{figure}[htbp]
    \centering
    \begin{subfigure}[t]{\textwidth}
        \centering
        \includegraphics[width=0.95\textwidth]{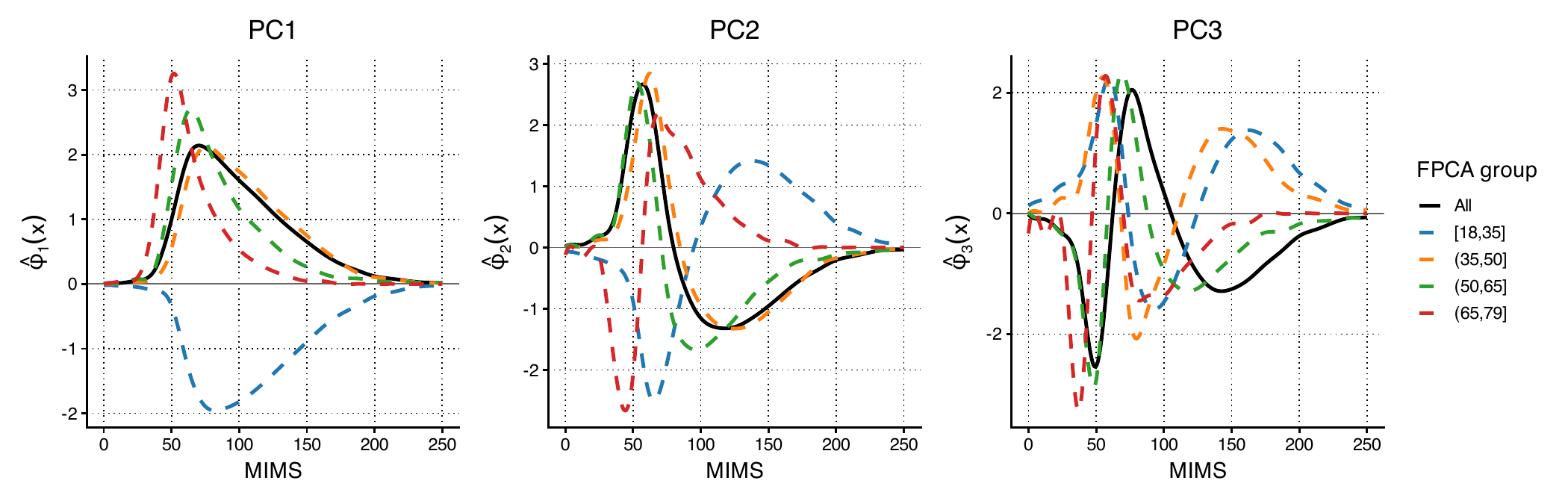}
        \caption{FPCA eigenfunctions stratified by age group}
    \end{subfigure}
    
    \begin{subfigure}[t]{\textwidth}
        \centering
        \includegraphics[width=\textwidth]{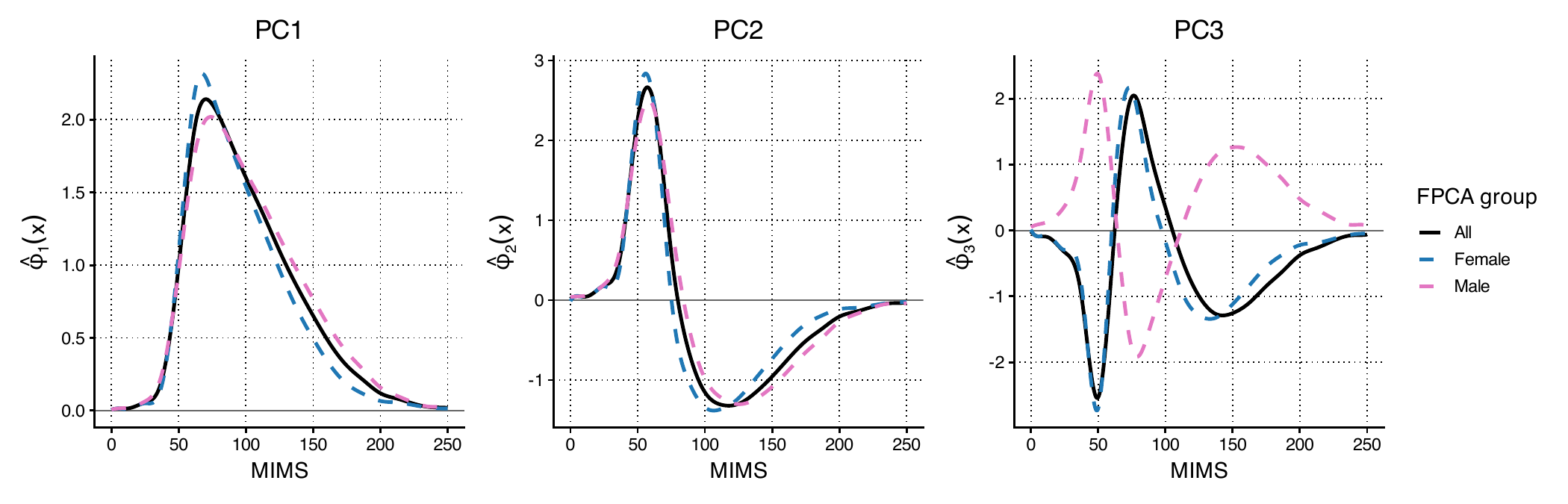}
        \caption{FPCA eigenfunctions stratified by gender}
    \end{subfigure}
    
    \begin{subfigure}[t]{\textwidth}
        \centering
           \includegraphics[width=0.95\textwidth]{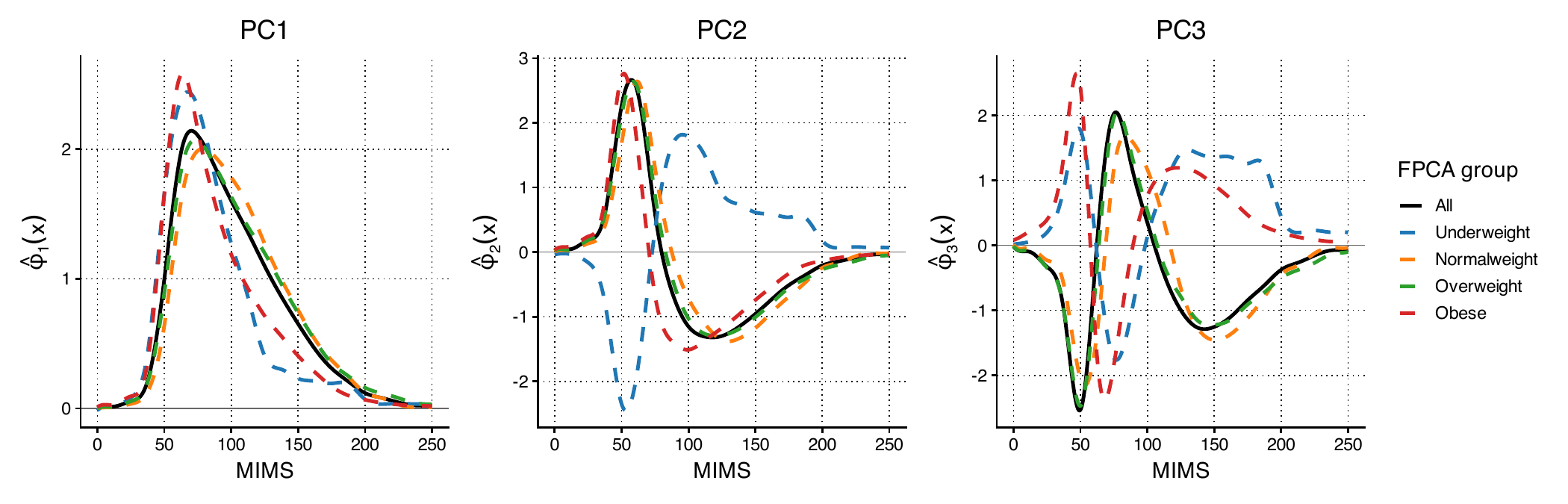}
        \caption{FPCA eigenfunctions stratified by obesity status}

    \end{subfigure}
    \begin{subfigure}[t]{\textwidth}
        \centering
        \includegraphics[width=0.95\textwidth]{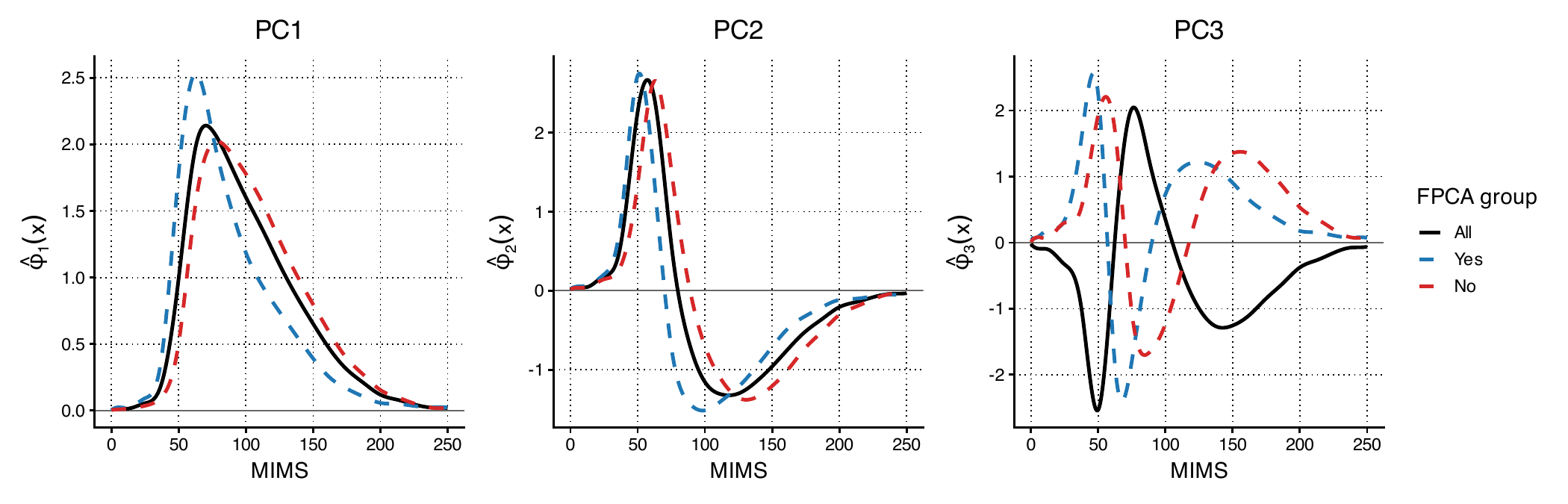}
        \caption{FPCA eigenfunctions stratified by hypertension status}
    \end{subfigure}
    
    \caption{First three eigen-functions using models \eqref{eq:fpca} and \eqref{eq:haz} based on MIMS.}

    \label{fig:fpca_eigenfunctions_mims}
\end{figure}

\newpage
\bibliographystyle{apalike}
\bibliography{cite}

\end{document}